\documentclass[%
 aip,
 amsmath,amssymb,
 reprint,
]{revtex4-1}

\usepackage{graphicx}
\usepackage{dcolumn}
\usepackage{bm}

\usepackage[utf8]{inputenc}
\usepackage[T1]{fontenc}
\usepackage{mathptmx}
\usepackage{etoolbox}
\usepackage{float}
\usepackage{color}
\usepackage{tikz}
\usepackage{amsmath}
\DeclareUnicodeCharacter{2212}{-}
\DeclareUnicodeCharacter{0301}{-}
\DeclareUnicodeCharacter{03B5}{-}
\makeatletter
\def\@email#1#2{%
 \endgroup
 \patchcmd{\titleblock@produce}
  {\frontmatter@RRAPformat}
  {\frontmatter@RRAPformat{\produce@RRAP{*#1\href{mailto:#2}{#2}}}\frontmatter@RRAPformat}
  {}{}
}%
\makeatother
\begin{document}

\preprint{AIP/123-QED}

\title{Heterogeneous ice nucleation on model substrates}

\author{M. Camarillo}
\affiliation{Departamento de Química Física, Facultad de Ciencias Químicas, Universidad Complutense de Madrid, 28040 Madrid, Spain}
\affiliation{Departamento de Ingeniería Química Industrial y del Medio Ambiente, Escuela T\'{e}cnica Superior de Ingenieros Industriales, Universidad Polit\'{e}cnica de Madrid, 28006, Madrid, Spain}

\author{J. Oller-Iscar}
\affiliation{Departamento de Ingeniería Química Industrial y del Medio Ambiente, Escuela T\'{e}cnica Superior de Ingenieros Industriales, Universidad Polit\'{e}cnica de Madrid, 28006, Madrid, Spain}

\author{M.M. Conde}
\affiliation{Departamento de Ingeniería Química Industrial y del Medio Ambiente, Escuela T\'{e}cnica Superior de Ingenieros Industriales, Universidad Polit\'{e}cnica de Madrid, 28006, Madrid, Spain}

\author{Jorge Ramírez}
\affiliation{Departamento de Ingeniería Química Industrial y del Medio Ambiente, Escuela T\'{e}cnica Superior de Ingenieros Industriales, Universidad Polit\'{e}cnica de Madrid, 28006, Madrid, Spain}

\author{E. Sanz*}
\affiliation{Departamento de Química Física, Facultad de Ciencias Químicas, Universidad Complutense de Madrid, 28040 Madrid, Spain}
\email{esa01@ucm.es}

\date{\today}

\begin{abstract}
Ice nucleation is greatly important in areas as diverse as climate change, cryobiology, geology or food industry. Predicting the ability of a substrate to induce the nucleation of ice from supercooled water is a difficult problem. Here, we use molecular simulations to analyse how the ice nucleating ability is affected by the substrate lattice structure and orientation. We focus on different model lattices: simple cubic, body centred cubic and face centred cubic, and assess their ability to induce ice nucleation by calculating nucleation rates. Several orientations are studied for the case of the face centred cubic lattice. Curiously, a hexagonal symmetry does not guarantee a better ice nucleating ability. By comparing the body centred cubic and the cubic lattices we determined that there is a significant role of the underlying crystal plane(s) on ice nucleation. The structure of the liquid layer adjacent to the substrate reveals that more efficient nucleants induce a more structured liquid.
The most efficient substrates present a strong sensitivity of their ice nucleating ability to the lattice parameters. Introducing a novel methodological approach, we use Classical Nucleation Theory to estimate the contact angle of the ice nucleus on the studied substrates from the calculated nucleation rates. The method also provides the nucleation free energy barrier height, the kinetic pre-factor and the critical cluster size. The latter is in agreement with the nucleus size obtained through a microscopic analysis of the nucleation trajectories, which supports the validity of Classical Nucleation Theory down to small critical clusters.
\end{abstract}

\maketitle

\section{\label{sec:intro}Introduction}

Ice nucleation plays a critical role in a wide range of natural and technological processes, from cloud formation and precipitation in the atmosphere to cryopreservation, food storage, and aircraft safety. In the atmosphere, the initiation of ice crystals governs the microphysical evolution of clouds, influencing climate feedback mechanisms and the global water cycle. Despite its importance, the fundamental mechanisms that govern ice nucleation —whether homogeneous or heterogeneous— remain only partially understood due to the complexity and stochastic nature of the process. Understanding how and where ice forms is essential for improving weather prediction models, designing antifreeze materials, and controlling ice formation in industrial applications. As such, advancing our knowledge of ice nucleation at the molecular level is a key challenge  \cite{laaksonen2021nucleation,hoose2012heterogeneous,maeda2021brief}.

In recent years, a fair agreement has been reached between experiments and molecular simulation studies regarding the rate 
of homogeneous ice nucleation \cite{espinosa2018homogeneous,niu2019temperature}, lending confidence to 
the theoretical frameworks that underpin water modelling and Classical Nucleation Theory (CNT) \cite{gibbsCNT1,gibbsCNT2,volmerweber1926}. This convergence of approaches highlights the potential of computational methods to provide atomistic insight into a process that remains experimentally challenging to probe directly.

While homogeneous nucleation provides a useful theoretical framework, heterogeneous nucleation is widely recognized as the dominant mechanism for ice formation under atmospheric and most natural conditions \cite{sanz2013homogeneous,hoose2012heterogeneous,maeda2021brief}. This process occurs in the presence of foreign substances—ice nucleating particles —which catalyze ice formation at significantly higher temperatures than required for homogeneous nucleation. However, heterogeneous nucleation is inherently more complex due to the vast diversity of nucleating agents \cite{murray2012ice,kanji2017overview,hoose2010classical,harrison2016not}, which range from mineral dust and soot to biological structures and engineered surfaces. Each of these materials presents different surface chemistries, structures, and hydrophilicities, making it challenging to generalize their nucleating efficiencies or derive universal principles. Despite substantial progress through experimental \cite{murray2012ice,hoose2012heterogeneous,atkinson2013importance,zhang2018control,whale2015ice,sosso2022role}, simulation \cite{lupi2016pre,lupi2014heterogeneous,fraux2014note,soni2021microscopic,zielke2015molecular,zielke2016simulations,soni2021unraveling,pedevilla2018heterogeneous,cox2013microscopic,sosso2016ice,glatz2016surface,sosso2018unravelling,sosso2022role,pedevilla2016can} and theoretical \cite{hoose2010classical,knopf2013water,cabriolu2015ice,khvorostyanov2000new,zobrist2008heterogeneous,liu2005ice,chen2008parameterizing,barahona2009parameterizing} studies, a unified molecular-level understanding of heterogeneous ice nucleation remains elusive, in contrast to the relatively consistent picture that has emerged for homogeneous nucleation. Developing a comprehensive framework for heterogeneous nucleation is a critical step toward accurately modelling ice formation in atmospheric and applied systems.

Given the complex interplay between structural and chemical factors influencing ice nucleation \cite{bi2017enhanced,fitzner2015many,glatz2018heterogeneous,cox2015molecularI,cox2015molecularII,valeriani2022deep,soni2021microscopic,reinhardt2014effects,li2017roles,zhang2018control,lupi2014does,lu2021effect,cox2013microscopic,fitzner2020predicting}, we recently chose to  
focus on the role of lattice mismatch \cite{camarillo2024effect}. To this end, we employed model substrates of stretched/compressed ice (composed of water molecules). This approach allowed us to directly assess the impact of lattice mismatch on nucleation behavior. Our results revealed that for each percent increase in lattice mismatch, the temperature at which ice nucleation occurs decreases by approximately 4 K. This finding underscores the sensitivity of heterogeneous ice nucleation to structural alignment and highlights the importance of epitaxial compatibility in promoting efficient ice formation.

Building on our previous results for ice Ih-like substrates \cite{camarillo2024effect}, we perform numerical simulations of ice nucleation on substrates composed of water molecules arranged in various crystalline lattices, including face-centered cubic (fcc), body-centered cubic (bcc), and simple cubic structures. By systematically varying the lattice geometry and orientation while maintaining the inter-molecular interaction potential, we are able to directly assess how the structural arrangement of the substrate influences its ability to promote ice nucleation.  

The strategy of using a substrate composed of molecules of the same type as the liquid
has already been employed to answer fundamental questions regarding the influence of the
substrate structure on its nucleating ability for Lennard-Jones particles \cite{mithen2014computer,mithen2014epitaxial}  and for water \cite{reinhardt2014effects,camarillo2024effect}.
The approach of using generic substrates contrasts with that followed in simulation works that use models of real substrates for ice nucleation \cite{kiselev2017active,fraux2014note,soni2021microscopic,zielke2015molecular,zielke2016simulations,soni2021unraveling,pedevilla2018heterogeneous,cox2013microscopic,sosso2016ice,glatz2016surface,sosso2018unravelling,sosso2022role,pedevilla2016can}, which are aimed at providing quantitative predictions (and require
special care in assessing the performance of the employed model potential). 
Our goal, instead, is to use simple generic 
substrates and compare their relative ice nucleating ability in order to learn something about
the effect of substrate structure on ice nucleating ability. 

 Moreover, by comparing the nucleation rates obtained in this work with those from our previous investigations on homogeneous ice nucleation  \cite{espinosa2014homogeneous,espinosa2016time,espinosa2016interfacial}, we are able to estimate the contact angle between the ice nucleus and the substrate. This approach, which is based in the CNT extension to heterogeneous nucleation \cite{volmer1929keimbildung,turnbull1950kinetics}, provides a novel method for estimating contact angles in simulations of heterogeneous nucleation, without the need to fit the interfacial nucleus shape,
which is not a straightforward task \cite{vuckovac2019uncertainties,jiang2019recent,santiso2013calculation,wang2022contactanglecalculator}, particularly for solid-fluid interfaces. Contact angles can also be obtained by determining all interfacial free energies involved in heterogeneous nucleation through complex
thermodynamic integration 
schemes \cite{reinhardt2014effects}.
Instead, our method relies solely on nucleation rate calculations and on CNT, offering a consistent and thermodynamically grounded means to quantify the wetting behavior of a substrate with respect to ice.

%

\section{\label{sec:meth}Methodology}
\subsection{Simulation details and set up}
We use the mW water model \cite{molinero2009water}, whose melting temperature is 273 K \cite{hudait2016free}.
We run our simulations with the Large-scale
Atomic/Molecular Massively Parallel Simulator (LAMMPS) Molecular Dynamics package \cite{plimpton1995fast} in the 
NVT ensemble. Temperature is kept constant with the Nosé-Hoover 
thermostat \cite{nose1984unified,hoover1985canonical}.
We integrate the equations
of motion using the velocity-Verlet integrator with a
3 fs time step. 

We study heterogeneous ice nucleation on substrates composed of water molecules that interact with
molecules belonging to the liquid via the mW potential.  
By keeping the same interactions between the liquid and the substrate molecules we
focus on the role of the substrate structure rather than on that of the interactions. 

\begin{figure}[H]
    \centering
    \includegraphics[clip,scale=0.42,angle=0.0]{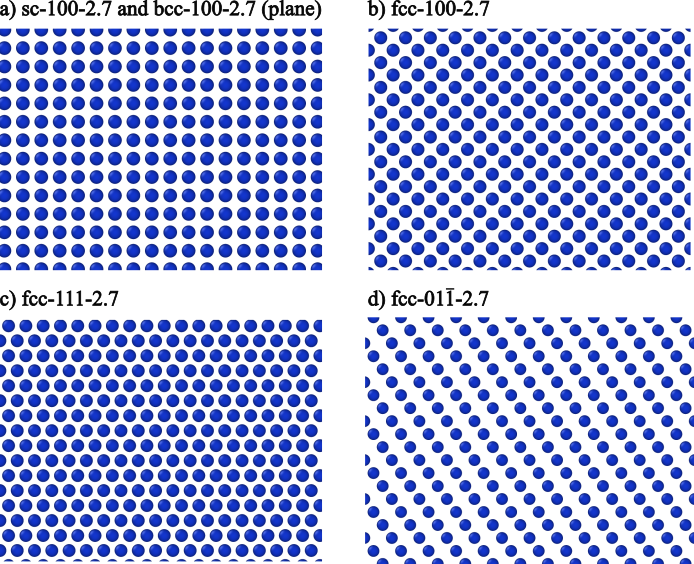}
    \caption{Plane exposed to the liquid \textcolor{black}{for the different substrates under study. In (a) the 100 orientation 
    of the sc and bcc lattices. In (b), (c) and (d) the 100, 111 and 01$\bar{1}$ planes of the
    fcc lattice, respectively.}
    }
    \label{fig:caras}
\end{figure}

We use substrates with face centred cubic (fcc), simple cubic (sc) and body centred cubic (bcc)
structures. In the case of fcc substrates, the 111, 100 and 01$\bar{1}$ planes are exposed
to the liquid to study the effect of lattice orientation on ice nucleation. 
In Fig. \ref{fig:caras} we show a snapshot of the different planes that were exposed to the liquid
(with a single-molecule thickness). 
Some of them are fully equivalent: the 100 planes of the fcc, sc and bcc lattices are 2D square
lattices identical to each other. However, as we will show later on, the ice nucleating ability
of these substrates is not  
the same given that underlying planes have some influence in the 
ability of a substrate to nucleate ice.

Different unit cell sizes are also investigated for each lattice type. 
In particular, we focus on the influence of nearest neighbor distance (nnd) 
on the ability of a given substrate to nucleate ice. 

We use the following code to identify a given substrate lattice and orientation: 
``lattice-plane-ndd (in \AA)''. For example, when exposing to the liquid the 100 orientation of an fcc lattice with 2.7 \AA~ nearest neighbor
distance we simply write: fcc-100-2.7. 
In all cases, except in the bcc-100 system, the nnd in the 3D lattice
coincides with that in the exposed plane. For the bcc-100 system, however, it is not equivalent to 
have a certain nnd in the exposed plane (given by the unit cell side) or in the 3D lattice
(given by half diagonal of the unit cell). To differentiate between these two cases we add the
letter ``l'', for  ``lattice'', or ``p'', for  ``plane'', at the end of the system code (e. g. 
bcc-100-2.7-l versus bcc-100-2.7-p).
We report in table 
\ref{tab:systemdetails} the details on the different substrates under consideration. 

\begin{figure}[H]
    \centering
    \includegraphics[width=0.53\linewidth, angle=90]{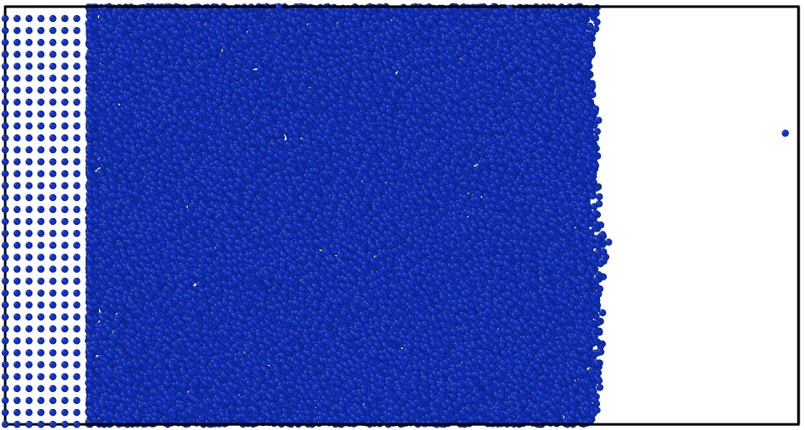}(a)
    \includegraphics[width=0.53\linewidth, angle=90]{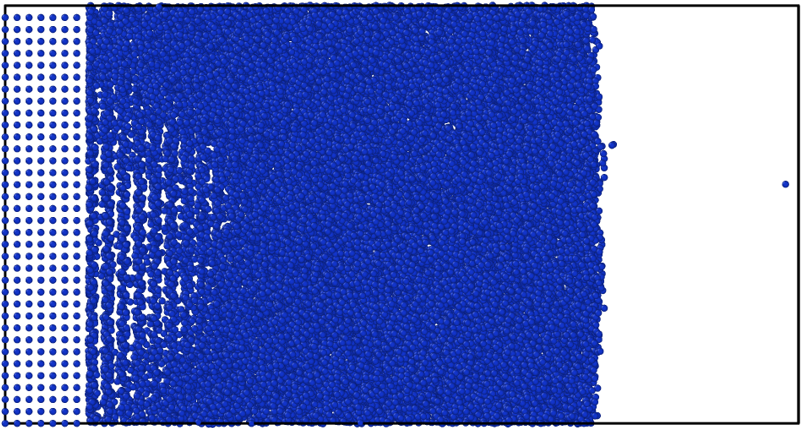}(b)
    \includegraphics[width=0.53\linewidth, angle=90]{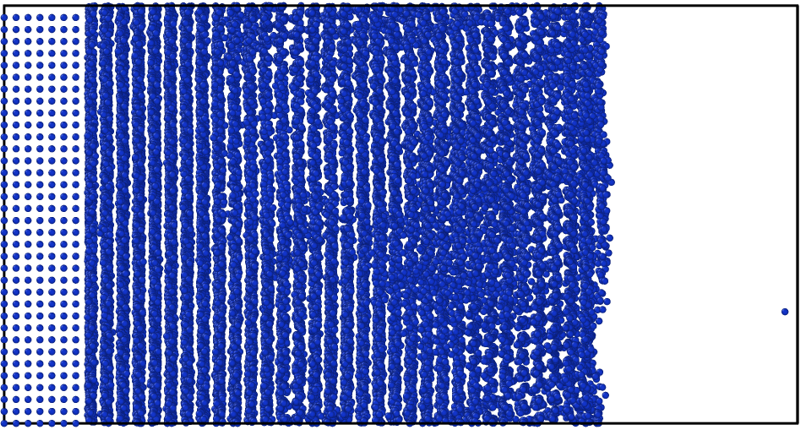}(c)
    \caption{a) Snapshot of a typical configuration of a liquid in contact with an sc-100-2.7 substrate. After simulating 15.1 ns at 221.5 K an ice nucleus appears (shown in (b)) and the whole 
    system quickly turns into ice (shown in (c)).}
    \label{fig:snapshot}
\end{figure}

Molecules in the substrate remain immobile in their lattice positions. 
The pressure of the liquid in contact with the substrate is $\sim$ 0 bar because the liquid
has a free interface. In Fig. \ref{fig:snapshot} (a) we show a typical snapshot 
illustrating our simulation setup. 

\begin{table}[htbp]

\scriptsize

\caption{Details on the systems employed in the present work.
    The substrate lattice, orientation and first neighbor distance are indicated
    with the following substrate code: "lattice type"-"miller indices of exposed plane"-"distance between first neighbors". In the case of the bcc lattice we also indicate whether the first neighbor distance
    refers to molecule pairs in the 2D mono-atomic width plane exposed to the liquid ("p") or to the 3D lattice ("l").  For the other lattices under consideration both distances coincide and
    there is no need for any specification. 
    We also report in the table the
    number of molecules in the substrate, the number of crystal planes, the 
    number of liquid molecules, the area of the substrate and 
    the dimensions of the simulation box. The x direction is perpendicular to the 
    substrate whereas the y and z directions are tangential to it.}

\centering

\begin{tabular*}{\linewidth}{@{\extracolsep{\fill}} lrrrrrrrr}

\hline

\textbf{Substrate} & \textbf{$N_\mathrm{substrate}$} & \textbf{$N_\mathrm{planes}$} & \textbf{$N_\mathrm{liq}$} & \textbf{A/nm$^2$} & \textbf{$L_x$ / nm} & \textbf{$L_y$ / nm} & \textbf{$L_z$ / nm} \\

\hline

sc-100-2.5 & 8575 & 7 & 34300 & 76.56 & 17.25 & 8.75 & 8.75 \\

sc-100-2.7 & 8575 & 7 & 34300 & 89.30 & 17.95 & 9.45 & 9.45 \\

sc-100-2.85 & 5400 & 6 & 18000 & 73.10 & 15.70 & 8.50 & 8.55 \\

sc-100-3.0 & 7350 & 6 & 35525 & 110.25 & 19.00 & 10.50 & 10.50 \\

sc-100-3.2 & 7350 & 6 & 35525 & 125.44 & 19.70 & 11.20 & 11.20 \\
\hline
bcc-100-2.5-l & 3600 & 9 & 8400 & 33.33 & 14.73 & 5.77 & 5.77 \\

bcc-100-2.7-l & 4000 & 10 & 8000 & 38.88 & 15.08 & 6.24 & 6.24 \\

bcc-100-2.7-p & 4400 & 11 & 7600 & 29.16 & 15.05 & 5.40 & 5.40 \\

bcc-100-3.0-l & 3600 & 9 & 8400 & 48.00 & 15.20 & 6.93 & 6.93 \\

bcc-100-3.2-l & 3600 & 9 & 8400 & 54.61 & 15.04 & 7.39 & 7.39 \\
\hline
fcc-100-2.5 & 7200 & 9 & 16800 & 50.00 & 15.30 & 7.07 & 7.07 \\

fcc-100-2.7 & 12168 & 9 & 28392 & 98.56 & 20.03 & 9.93 & 9.93 \\

fcc-100-2.85 & 7200 & 9 & 24000 & 64.98 & 16.05 & 8.06 & 8.06 \\

fcc-100-3.0 & 6400 & 8 & 17600 & 72.00 & 16.36 & 8.49 & 8.49 \\

fcc-100-3.2 & 5600 & 7 & 18400 & 81.92 & 16.79 & 9.05 & 9.05 \\
\hline
fcc-111-2.5 & 4500 & 5 & 16200 & 48.72 & 12.00 & 6.50 & 7.50 \\

fcc-111-2.7 & 4500 & 5 & 16200 & 56.82 & 10.00 & 7.02 & 8.10 \\

fcc-111-2.85 & 4500 & 5 & 16200 & 63.31 & 10.00 & 7.40 & 8.55 \\

fcc-111-3.0 & 4500 & 5 & 16200 & 70.15 & 10.00 & 7.79 & 9.00 \\

fcc-111-3.2 & 4500 & 5 & 16200 & 79.82 & 10.00 & 8.31 & 9.60 \\
\hline
fcc-01$\bar{1}$-2.5 & 9450 & 14 & 31050 & 59.66 & 22.50 & 6.50 & 9.19 \\

fcc-01$\bar{1}$-2.7 & 8775 & 13 & 31725 & 69.59 & 20.00 & 7.02 & 9.92 \\

fcc-01$\bar{1}$-2.85 & 15600 & 13 & 108000 & 137.85 & 34.03 & 9.87 & 13.96 \\

fcc-01$\bar{1}$-3.0 & 8100 & 12 & 32400 & 85.92 & 25.00 & 7.79 & 11.02 \\

fcc-01$\bar{1}$-3.2 & 8100 & 12 & 32400 & 97.75 & 25.60 & 8.31 & 11.76 \\

\hline
\label{tab:systemdetails}
\end{tabular*}

\end{table}

\subsection{Heterogeneous nucleation rate}
The ability of a substrate to nucleate ice is quantified by the
heterogeneous nucleation rate, $J_{het}$, which is the number
of ice nuclei that emerge per unit of time and area. 
To calculate $J_{het}$ we use:
\begin{equation}
    J_{het}=\frac{1}{t_nA}
    \label{eq:rate}
\end{equation}
where $A$ is the area of the substrate exposed to the liquid and $t_n$ is the average
time required for a nucleus to appear. 
For sufficiently low temperatures, the appearance of the nucleus occurs 
spontaneously and stochastically after some induction time. 
The average induction time over different independent trajectories
gives $t_n$ and, hence, $J_{het}$. In Fig. \ref{fig:snapshot} we show snapshots 
of the appearance of an ice cluster on an sc substrate (part (b)) 
followed by the crystallisation of the whole system (part (c)).

\subsection{Analysis}

\subsubsection{Structural analysis}
To try to understand the relative ice-nucleating efficiency of different substrates, we analyse the molecular structure across the interface. Specifically, we calculate the radial distribution function (rdf) for particles within thin slabs —each a few \AA~ thick— parallel to the interface, and compare these to the rdf of bulk ice. 

We also analyse the number of particles that belong to the biggest
crystalline cluster along the nucleation trajectory. A molecule 
is considered ``ice-like'' if its
$\bar{q}_6$  \cite{lechner2008accurate} is larger than a certain threshold tuned
with the mislabelling criterion as indicated in Ref. \cite{espinosa2016interfacial}.

\subsubsection{Contact angle}
\label{sec:contact}

\begin{figure}[H]
    \centering
    \includegraphics[clip,scale=0.22,angle=0.0]{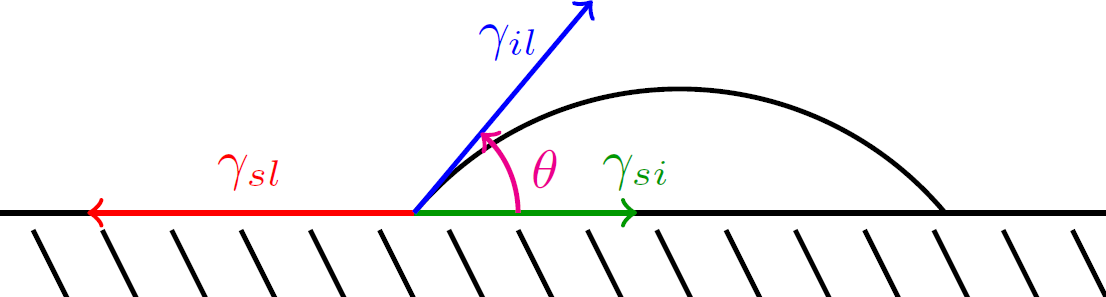}
    \caption{Contact angle of a nucleus on a substrate as a balance of the interfacial free energies involved.}
    \label{fig:contact-angle}
\end{figure}

We employ CNT \cite{gibbsCNT1,gibbsCNT2,volmerweber1926,becker-doring} 
extended to heterogeneous nucleation \cite{volmer1929keimbildung,turnbull1950kinetics}
to estimate the contact angle, $\theta$, of the ice nucleus on each type of substrate. 
The contact angle 
depends
on a balance of the substrate-ice ($s-i$), ice-liquid ($i-l$) and substrate-liquid ($s-l$) interfacial free energies given by Young's equation \cite{young1805iii}:
\begin{equation}
    \gamma_{s-l}=\gamma_{i-l}\cos(\theta)+\gamma_{s-i}.
\end{equation}
See the sketch in Fig. \ref{fig:contact-angle} for a schematic representation of the contact
angle of an ice nucleus on a substrate as a balance of the three interfacial free energies. 

The nucleus volume and surface area can be expressed in terms of those of a sphere of the same 
radius of curvature times a fraction that depends on $\theta$.
Specifically, the volume of the cap-shaped heterogeneous nucleus is given by, 
\begin{equation}
V_{\text{het}} = V_{\text{hom}} f(\theta),
\label{eq:vhomvhet}
\end{equation}
with
\begin{equation}
f(\theta) = \frac{(2 + \cos\theta)(1 - \cos\theta)^2}{4}.
\end{equation}
$V_{hom}$ in Eq. \ref{eq:vhomvhet} is the volume of a 
spherical nucleus that has the same radius of curvature 
as the spherical cap. 
For $\theta=180^{\text{o}}$ (no wetting) the nucleus volume takes the value of that of the full sphere and 
for $\theta=0$ (full wetting) it becomes 0.
On the other hand, the nucleus surface area in heterogeneous nucleation is related
to that in homogeneous nucleation by:
\begin{equation}
    A_{het}=A_{hom} h(\theta)
\end{equation}
where 
\begin{equation}
    h(\theta)=(1-\cos(\theta))/2.
\end{equation}

By developing the CNT formalism with these new expressions for the volume and the area
of the nucleus \cite{volmer1929keimbildung}, one finds that the free energy barriers for the homogeneous and the heterogenous
nucleation mechanisms are related by:
\begin{equation}
    \Delta G^c_{het}=\Delta G^c_{hom} f(\theta).
    \label{eq:deltaGhomhet}
\end{equation}
For $\theta=180^{\text{o}}$ there is no wetting and homogeneous and heterogeneous nucleation 
have the same barrier whereas for $\theta=0^{\text{o}}$
there is no nucleation barrier because the substrate is fully wetted by the nucleating phase. 
Note that homogeneous nucleation can be seen as a particular case of heterogeneous
nucleation for $\theta=180^{\text{o}}$.

Our aim is to estimate the contact angle using Eq. \ref{eq:deltaGhomhet}.
The homogeneous nucleation barrier is known from our previous work \cite{espinosa2016time,espinosa2014homogeneous,espinosa2016interfacial}. 
We can straightforwardly compare homogeneous and heterogeneous nucleation 
because we are using a short-range potential and we do not have to worry about 
long-range corrections \cite{atherton2022can}.

We obtain estimates of $\Delta G_{het}^c$ through the nucleation rate, $J_{het}$, calculated in simulations of spontaneous nucleation via Eq. 
\ref{eq:rate}.
To obtain $\Delta G_{het}^c$ from $J_{het}$
we assume the CNT formalism for the heterogeneous rate:
\begin{equation}
    J_{het}=\rho_s f^+_{het} Z_{het} \exp(-\Delta G^c_{het}/(k_BT))
    \label{eq:jhetcnt}
\end{equation}
where $\rho_s$ is the molecular surface density of the substrate, $f^+_{het}$ is the 
attachment rate of molecules to
the critical cluster
and $Z_{het}$ is the Zeldovich factor. 
$\rho_s$ can be easily obtained by counting the number of molecules per unit area in the exposed
plane of the substrate.
$f^+_{het}$ is the attachment frequency of molecules to the critical cluster, which
is proportional to the  contact area between the nucleus and the liquid. 
Such proportionality implies that we can obtain the attachment
rate of particles to the heterogeneous nucleus from that to the
homogeneous nucleus as:
\begin{equation}
    f^{+}_{het}=f^{+}_{hom} h(\theta).
\label{eq:f+}
\end{equation}
We take $f^+_{hom}$ from our previous studies of homogeneous nucleation \cite{espinosa2016interfacial,espinosa2016time}.

The Zeldovich factor for homogeneous nucleation is 
given by:
\begin{equation}
    Z_{hom}=\sqrt{\frac{\Delta \mu}{6\pi k_BT N_{c,hom}}}
\end{equation}
where $N_{c,hom}$ is the number of molecules in the critical cluster and 
$\Delta \mu$ is the chemical potential difference between liquid and ice.
Taking into account that $N_{c,het}=f(\theta)N_{c,hom}$ because the
number of molecules and the volume are proportional to each other and that
$\Delta \mu$ (the chemical potential difference between ice and the supercooled liquid) is the same 
in homogeneous and heterogeneous nucleation one has:
\begin{equation}
    Z_{het}=\frac{Z_{hom}}{\sqrt{f(\theta)}}
    \label{eq:Z}
\end{equation}
The Zeldovich factor is higher in heterogeneous nucleation, which is reasonable because the
barrier is lower and its curvature is higher
(recall that the Zeldovich factor is proportional to the square root of the barrier curvature). Consequently, the fraction of critical nuclei that
eventually grow in heterogeneous nucleation is larger in heterogeneous than in homogeneous nucleation. 

Knowing the kinetic pre-factor ($\rho_s f^+_{het} Z_{het}$), and the 
nucleation rate, $J_{het}$, we obtain $\Delta G_{het}^c$
via Eq. \ref{eq:jhetcnt}. Then, using Eq. \ref{eq:deltaGhomhet}, 
we obtain the contact angle $\theta$.

In summary, we compute $J_{het}$  and by using CNT and comparing with our previous work 
on homogeneous nucleation we obtain the nucleation barrier for heterogeneous nucleation,
the kinetic pre-factor, and the contact angle. Moreover, since we have the number of particles
in the critical cluster for homogeneous nucleation we can estimate the corresponding number
for the heterogeneous cluster by multiplying by $f(\theta)$. 

\section{Results}

\subsection{Nucleation rates and temperatures}
\label{sec:rate_Tn}
We compute the heterogeneous ice nucleation rate 
via Eq. \ref{eq:rate}, which assumes that only one nucleus
appears and grows after some induction period. 
There is only a narrow temperature window where Eq. \ref{eq:rate} is applicable. 
At high temperatures the nucleation rate is too low and there is no nucleation 
in the duration of our simulations (lasting a few tens of ns), whereas
at low temperatures there are multiple nucleation events straight away from 
the beginning of the simulation. 
For each type of substrate listed on Table \ref{tab:systemdetails} we have to 
find the temperature range for which there is nucleation of a single nucleus.

\begin{figure}[H]
    \centering
    \includegraphics[clip,scale=0.22,angle=0.0]{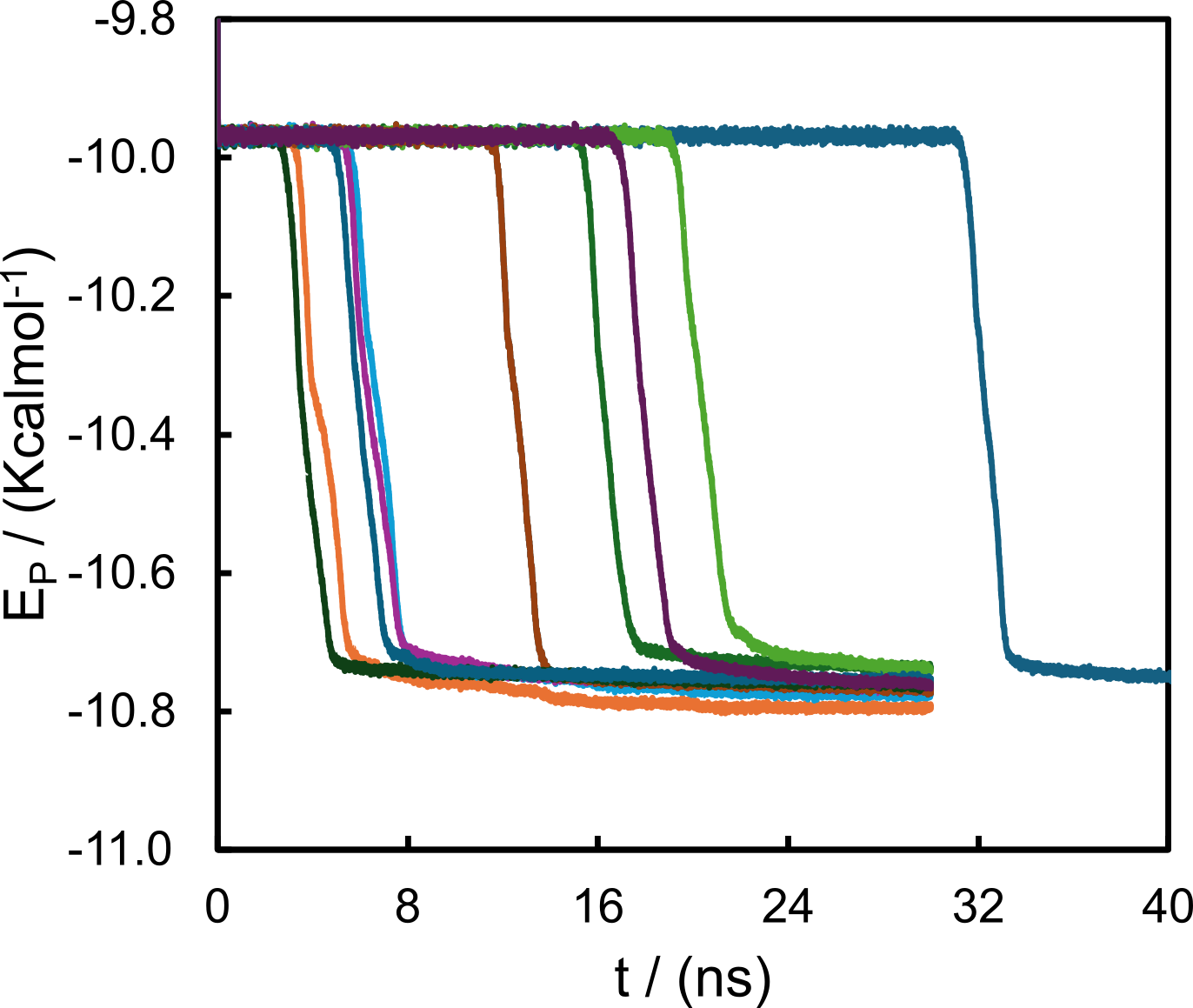}
    \caption{Potential energy vs time for 10 different trajectories at 221.5 K on an sc-100-2.7 substrate. The sudden steps are due to nucleation and growth of ice.}
    \label{fig:stochasticnucleation}
\end{figure}

The potential energy versus time in the nucleation regime we are interested in looks like
the curves shown in Fig. \ref{fig:stochasticnucleation}: first, there is an initial plateau corresponding
to the induction period where the liquid remains metastable (undercooled with respect to ice); then,
a sharp decrease of the potential energy is observed due to the emergence of an ice nucleus that quickly grows; 
finally, the curves plateau again due to the crystallization of the whole liquid.
The stochastic nature of nucleation is reflected in the fact that the nucleation time
--identified with
the time at which the potential energy suddenly drops--
largely varies from one trajectory to another. The average nucleation time, $t_n$, 
is used to compute, alongside the substrate area in contact with the liquid, $A$, the nucleation rate
via Eq. \ref{eq:rate}.

\begin{figure}[H]
    \centering
    \includegraphics[clip,scale=0.19,angle=0.0]{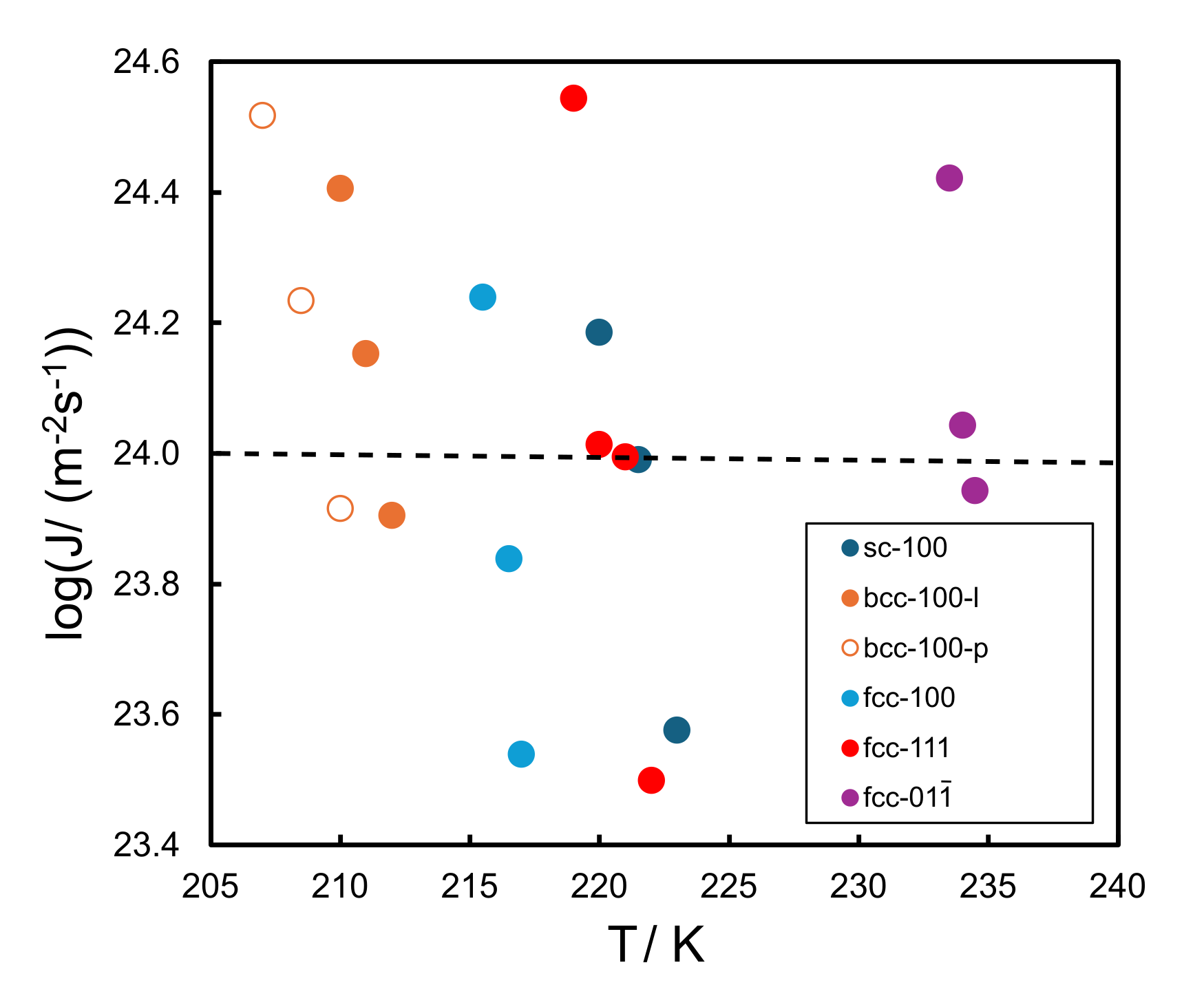}
    \caption{Decimal logarithm of the nucleation rate versus 
    temperature for different substrates having an nnd of 2.7 \AA (see legend).}
    \label{fig:rates}
\end{figure}

In Fig. \ref{fig:rates} we plot the nucleation rate versus 
the temperature for different substrates
having an nnd of 2.7 \AA. We obtain, in all cases, rates of the order of 10$^{23}$-10$^{24}$
m$^{-2}$ s$^{-1}$. It is not a coincidence that we always obtain rates of the same 
order of magnitude: the nucleation rate is determined by the length and time scales
accessible in our simulations. For simulation times of a few tens of ns and 
substrate sides of the order of 10 nm we obtain rates of 
1/(10$\cdot$10$^{-9}$s$\cdot$100$^{-18}$m$^{2}$)=10$^{24}$s$^{-1}$m$^{-2}$.
Therefore, what differs from one substrate to another is not the
rate, which is determined by the simulation set up, but the temperature
at which the rate accessible by our simulations is reached.
By interpolation in Fig. \ref{fig:rates} we find $T_N$, the temperature for which 
$\log[J/($m$^{-2}$s$^{-1}$$)]$=24, which is an arbitrary value within the range of 
accessible nucleation rates. 
The nucleation temperature, $T_N$, serves to quantify and compare the ice nucleating ability
of different substrates. 
Substrates with a high $T_N$ have a high ability to nucleate ice because the
accessible rate is reached at higher temperatures. 
In Fig. \ref{fig:rates} it becomes evident that the nucleating
ability of different substrates can be ranked, from the best to the
worst nucleant, as: fcc-01$\bar{1}$, fcc-111,  sc-100, fcc-100 and bcc-100
(fcc-111 and sc-100 have nearly the same $T_N$).
The nucleation temperature of the different substrates examined
in this work is
reported in Table \ref{tab:rate-results}.

\begin{figure}[H]
    \centering
    \includegraphics[clip,scale=0.20,angle=0.0]{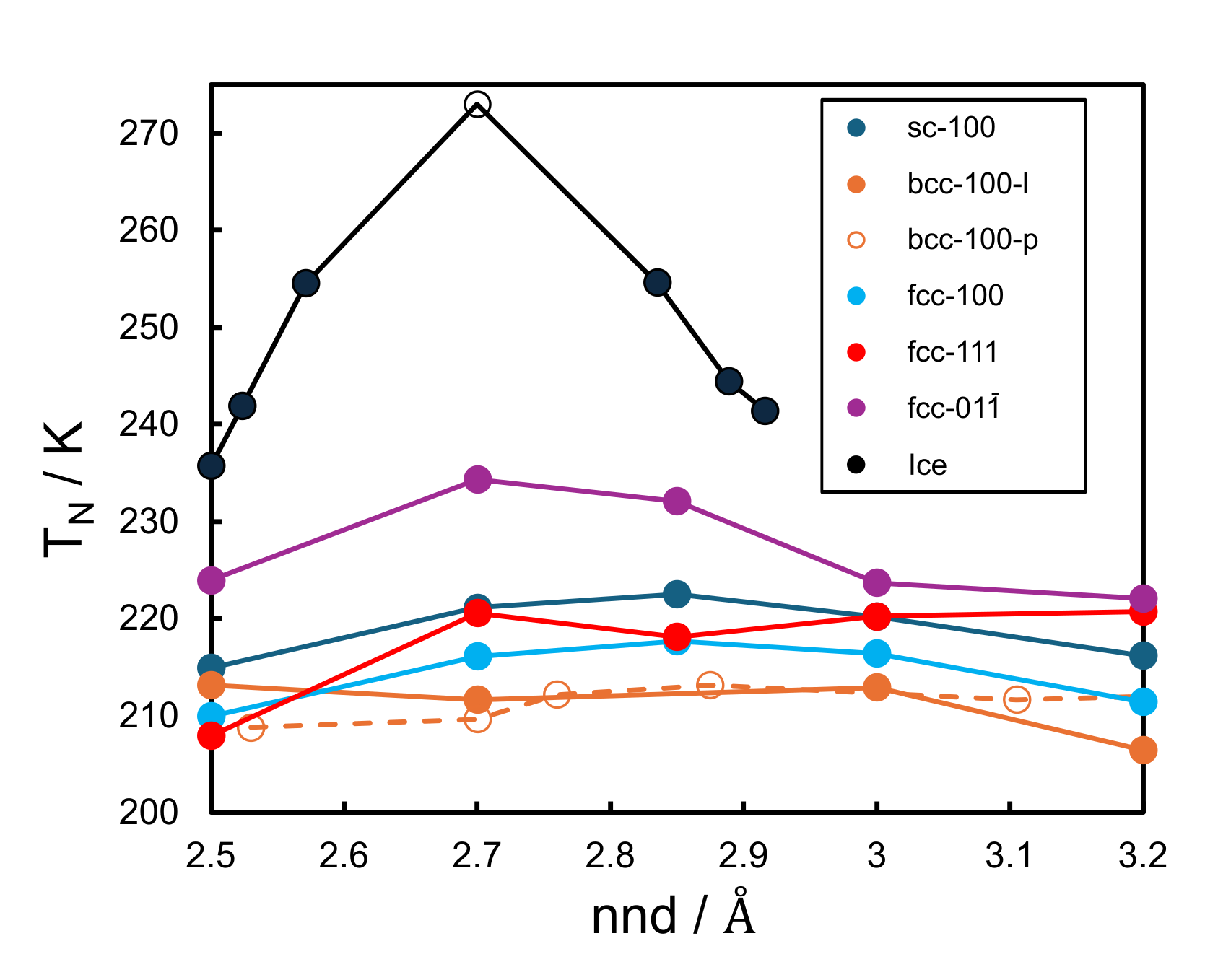}
    \caption{Nucleation temperature versus nearest neighbor distance
    for all the lattice orientations studied in this work (see legend).}
    \label{fig:tnvsd}
\end{figure}

In Fig. \ref{fig:tnvsd} we plot $T_N$ for all substrates studied in this work
as a function of the distance between first neighbours. 
Errors are of the order of the symbol size and lines are eye guides. 
The bcc lattice has two different curves, one corresponding to 
nnd in the lattice (solid) and the other corresponding to nnd in 
the exposed plane (dashed). Their ability to induce nucleation is similar
because $T_N$ does not change much with the nnd
for the bcc substrate. 

\begin{table}[htbp]

\scriptsize

\caption{Nucleation parameters for different substrates.}

\centering

\begin{tabular*}{\linewidth}{@{\extracolsep{\fill}} lcccccccccccc}

\hline

\textbf{Substrate} & $T_N/K$ & $\rho_s/($nm$^{-2})$ & $f^+/(10^{12}s^{-1})$& $Z\cdot10^2$ & $\frac{\Delta G_{\text{het}}}{k_BT}$ & $h(\theta)$ & $f(\theta)$ & $\theta$& $N_c$ \\

\hline

sc-100-2.5 & 214.9 & 16.00 & 5.7 & 2.5 & 14.6 & 0.44 & 0.42 & 83 & 51 \\

sc-100-2.7 & 221.1 & 13.71 & 7.6 & 2.2 & 14.6 & 0.38 & 0.32 & 76 & 57 \\

sc-100-2.85 & 222.5 & 12.31 & 8.1 & 2.1 & 14.6 & 0.36 & 0.30 & 74 & 59 \\

sc-100-3.0 & 220.5 & 11.11 & 7.1 & 2.2 & 14.4 & 0.39 & 0.33 & 77 & 55 \\

sc-100-3.2 & 216.2 & 9.77 & 5.9 & 2.5 & 14.2 & 0.42 & 0.39 & 81 & 50 \\
\hline
bcc-100-2.5-l & 213.1 & 12.00 & 5.1 & 2.6 & 14.3 & 0.47 & 0.44 & 86 & 48 \\

bcc-100-2.7-l & 210.4 & 10.29 & 4.4 & 2.8 & 14.0 & 0.48 & 0.49 & 88 & 44 \\

bcc-100-2.7-p & 208.5 & 13.71 & 4.3 & 2.8 & 14.3 & 0.52 & 0.51 & 92 & 44 \\

bcc-100-3.0-l & 213.4 & 8.33 & 5.0 & 2.6 & 13.9 & 0.47 & 0.43 & 86 & 46 \\

bcc-100-3.2-l & 206.4 & 7.32 & 3.7 & 2.9 & 13.6 & 0.54 & 0.56 & 95 & 40 \\
\hline
fcc-100-2.5 & 209.9 & 16.00 & 4.5 & 2.7 & 14.5 & 0.52 & 0.50 & 92 & 46 \\

fcc-100-2.7 & 216.1 & 13.71 & 6.0 & 2.4 & 14.5 & 0.44 & 0.40 & 83 & 51 \\

fcc-100-2.85 & 217.6 & 12.31 & 6.4 & 2.4 & 14.4 & 0.44 & 0.37 & 83 & 53 \\

fcc-100-3.0 & 216.9 & 11.11 & 6.0 & 2.4 & 14.3 & 0.43 & 0.38 & 82 & 51 \\

fcc-100-3.2 & 211.4 & 9.76 & 4.7 & 2.7 & 14.0 & 0.49 & 0.49 & 89 & 48 \\
\hline
fcc-111-2.5 & 207.9 & 18.79 & 4.1 & 2.8 & 14.6 & 0.53 & 0.54 & 93 & 44 \\

fcc-111-2.7 & 220.5 & 16.04 & 7.4 & 2.1 & 14.8 & 0.40 & 0.33 & 78 & 57 \\

fcc-111-2.85 & 218.2 & 7.19 & 6.4 & 2.4 & 13.9 & 0.40 & 0.35 & 79 & 54 \\

fcc-111-3.0 & 220.5 & 13.06 & 7.5 & 2.2 & 14.6 & 0.40 & 0.32 & 78 & 57 \\

fcc-111-3.2 & 220.7 & 11.05 & 7.3 & 2.3 & 14.4 & 0.40 & 0.32 & 79 & 56 \\
\hline
fcc-01$\bar{1}$-2.5 & 223.9 & 11.31 & 8.6 & 2.0 & 14.5 & 0.34 & 0.28 & 71 & 61 \\

fcc-01$\bar{1}$-2.7 & 234.3 & 9.70 & 14 & 1.6 & 14.6 & 0.27 & 0.17 & 62 & 78 \\

fcc-01$\bar{1}$-2.85 & 232.1 & 8.71 & 12 & 1.7 & 14.4 & 0.28 & 0.19 & 64 & 74 \\

fcc-01$\bar{1}$-3.0 & 224.6 & 7.86 & 8.3 & 2.1 & 14.1 & 0.35 & 0.28 & 73 & 58 \\

fcc-01$\bar{1}$-3.2 & 222.0 & 6.91 & 7.7 & 2.2 & 14.0 & 0.36 & 0.29 & 74 & 56 \\
\hline
ice-pII-2.835 & 254.1 & 5.63 & 38 & 0.7 & 14.3 & 0.11 & 0.03 & 39 & 181 \\

ice-pII-2.889 & 243.7 & 5.41 & 21 & 1.2 & 14.1 & 0.19 & 0.09 & 51 & 114 \\

ice-pII-2.916 & 240.7 & 5.32 & 18 & 1.3 & 14.1 & 0.21 & 0.11 & 54 & 102 \\ 
\hline
ice-pII-2.57 & 254.5 & 8.02 & 40 & 0.7 & 14.6 & 0.11 & 0.03 & 39 & 174 \\

ice-pII-2.52 & 241.9 & 8.37 & 20 & 1.2 & 14.6 & 0.20 & 0.10 & 53 & 100 \\

ice-pII-2.50 & 235.7 & 8.55 & 15 & 1.5 & 14.5 & 0.24 & 0.15 & 59 & 81 \\

\hline
\label{tab:rate-results}
\end{tabular*}

\end{table}

We include in the figure, in black, results for a hexagonal ice Ih substrate 
 exposing the secondary prismatic face (pII) from our previous work\cite{camarillo2024effect}. 
The ice-like substrate has the highest nucleation temperature, as expected, as it is the most effective at promoting ice nucleation thanks to its strong structural similarity to ice (it is in fact a stretched or compressed ice lattice). The open symbol in the ice curve is not really a nucleation temperature, 
but rather
the melting temperature of the model, which is assigned to an ice-like
substrate that is not deformed with respect to the relaxed ice lattice
(with  nnd = 2.7 \AA). 

Following the ice-like substrate, the fcc-01$\bar{1}$ 
system is the second most effective ice nucleant. 
It is perhaps surprising that this plane, that does not have a hexagonal symmetry, nucleates ice better than the 111 orientation that does have 
a hexagonal symmetry (see Fig. \ref{fig:caras}).
Perhaps, the fact that the hexagons present in the fcc-111 plane 
have a molecule in the centre --which is absent in the hexagons of 
the ice lattice-- justifies that the fcc-111 orientation is not an ice nucleant as
good as one might expect based on its hexagonal symmetry.
Moreover, there is another remarkable dissimilarity between fcc-111 and ice Ih basal planes: the hexagons 
are flat in the former and chair-like in the latter. 
The worse performance of the 111 orientation
raises a warning
that establishing an a priori correlation between substrate
structure and ice nucleating ability is not trivial. 
\textcolor{black}{Such a consideration is in agreement with the lack of correlation between 
similarity with the ice lattice and the ice nucleating ability of -OH patterned surfaces evidenced in Ref. \cite{pedevilladescriptorJCP2017}}.
At the end of the day, simulations or experiments must be done to 
characterise the nucleation rate and compare different substrates. 

The fcc-100 plane is the worst ice nucleant among all exposed orientations
of the fcc lattice. Its nucleation temperature typically lies between 5 K and 15 K below that of the 01$\bar{1}$ orientation. 
Note that 10 K difference in the nucleation temperature is a huge 
difference in the nucleation rate in view of the steep temperature
dependence of the nucleation rates shown in Fig. \ref{fig:rates}.

For a given nearest neighbour distance 
the, fcc-100, sc-100 and bcc-100-p substrates 
have the same atomic landscape in the outermost plane (a 2D cubic lattice 
with the same lattice parameter).
These substrates exhibit, however, systematic differences in their nucleation temperature (see Fig. \ref{fig:tnvsd}), which 
indicates that the plane(s) underneath the outermost plane 
strongly influence the 
ice nucleating ability of the substrate as a whole, event for a short-range water model as mW. 
In fact, the nucleation temperature of the best nucleant among these substrates
(sc-100) is about 10 K larger than that of the worst one (bcc-100).
Again, considering the steepness of $J_{het}(T)$ (Fig. \ref{fig:rates}), a 10 K difference in $T_N$
is a huge difference in $J_{het}$ for a given temperature.

Having compared the ice nucleating ability of different substrates among themselves, we now
focus on the slope of 
the $T_N$(nnd) curves shown in Fig. \ref{fig:tnvsd}.
Ice-like substrates (black curve) exhibit a strong slope
($\sim$ 20 K per 0.1 \AA)
when moving away from the nnd of the relaxed ice lattice (2.7 \AA).
The second best nucleant,  fcc-01$\bar{1}$, also shows a clear maximum 
near 2.7 \AA~ although the slope when moving away from the maximum
is smaller. 
It seems that the performance of the most effective ice-nucleating substrates --ice-like and fcc-01$\bar{1}$-- is more sensitive to variations in lattice density than that of less
efficient substrates.
The curves corresponding to the other substrates under consideration 
show a mild dependence with nnd.
Some of them are rather noisy and show no clear dependence with the nnd (bcc-100 and fcc-111 lattices) and some others (sc-100 and fcc-100) show a mild maximum close to 2.7 \AA. 

\textcolor{black}{
In summary, our results suggest that the better the substrate, the more sensitive its performance is
to the lattice parameters.
Although this is only an observation in view of our results and, as such, should be taken with care,
it seems quite reasonable
that more specific substrates are more sensitive to their structural details. 
If we focus on the case of ice-like substrates (black curve in Fig. \ref{fig:tnvsd}), it 
seems expected that as the ice
lattice expands or compresses with respect to the perfect (equilibrium) lattice the performance of these substrates
deteriorates from inducing ice formation at the highest possible temperature (the melting temperature) to doing
it at lower temperatures. In contrast, less specific substrates, show less 
sensitivity because they are poor nucleants in any case. A key parameter that may explain this behaviour is the substrate-ice interfacial
free energy, $\gamma_{s-i}$. Obviously, the smaller this parameter, the better the nucleant. For perfect ice-like
substrates (nnd=2.7\AA) this parameter is expected to be negligible, whereas it should significantly increase
as the ice-like substrate lattice
deviates from that of equilibrium ice. However, for non-specific substrates, $\gamma_{s-i}$ should be high 
for any lattice parameter, and not so sensitive to the lattice spacing. It would be very interesting in the
future to calculate $\gamma_{s-i}$ for different substrates and different lattice parameters 
in a systematic and comparative manner}.

\subsection{Kinetic pre-factor, free energy barrier and contact angle}

\begin{figure}[H]
    \centering
    \includegraphics[clip,scale=0.22,angle=0.0]{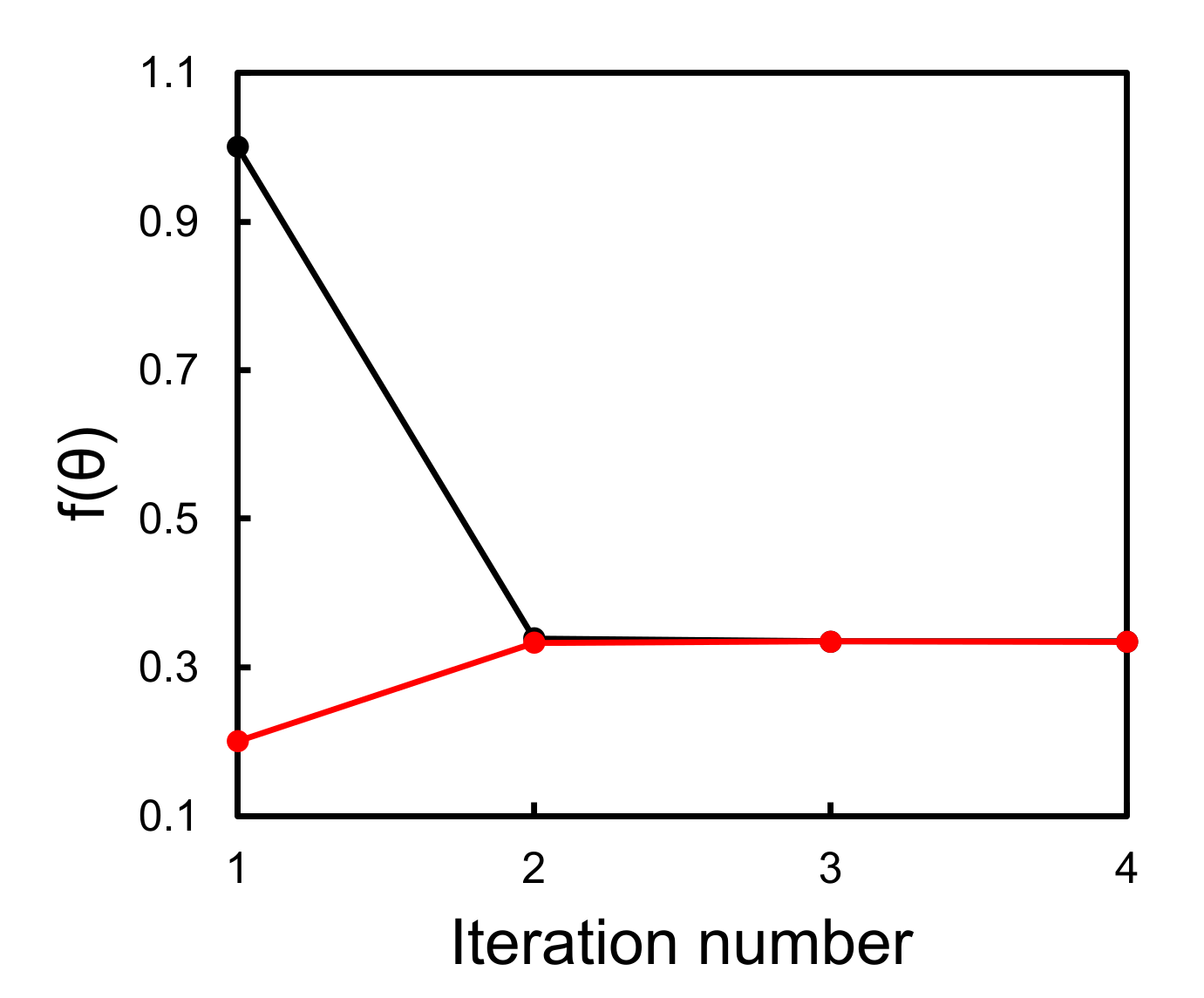}
        \caption{$f(\theta)$ along the iterative process to obtain $\Delta G^c_{het}, f^+_{het}$,
        $Z_{het}$ and $\theta$ for the sc-100-2.7 substrate.  Convergence is quickly reached for two
        different initial guess values for $f(\theta)$.}
    \label{fig:iterations}
\end{figure}

\begin{figure}[H]
    \centering
    \includegraphics[clip,scale=0.38,angle=0.0]{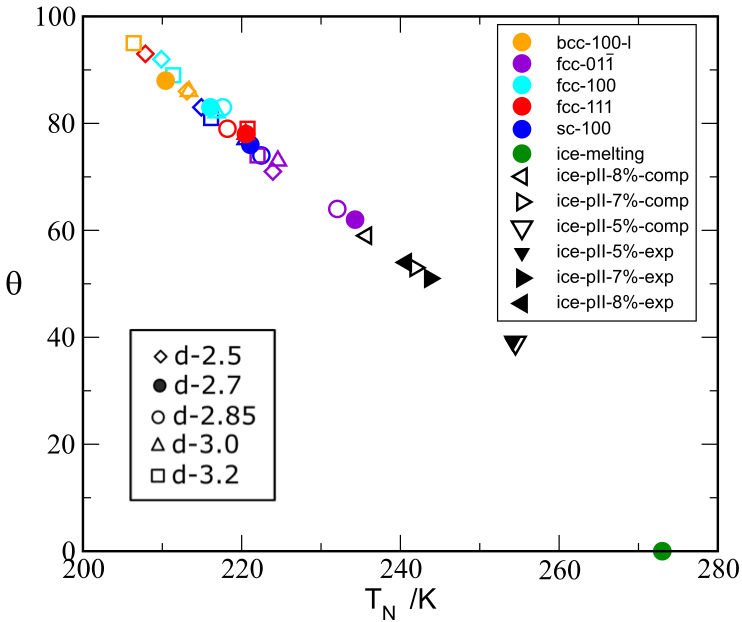} a)
    \includegraphics[clip,scale=0.19,angle=0.0]{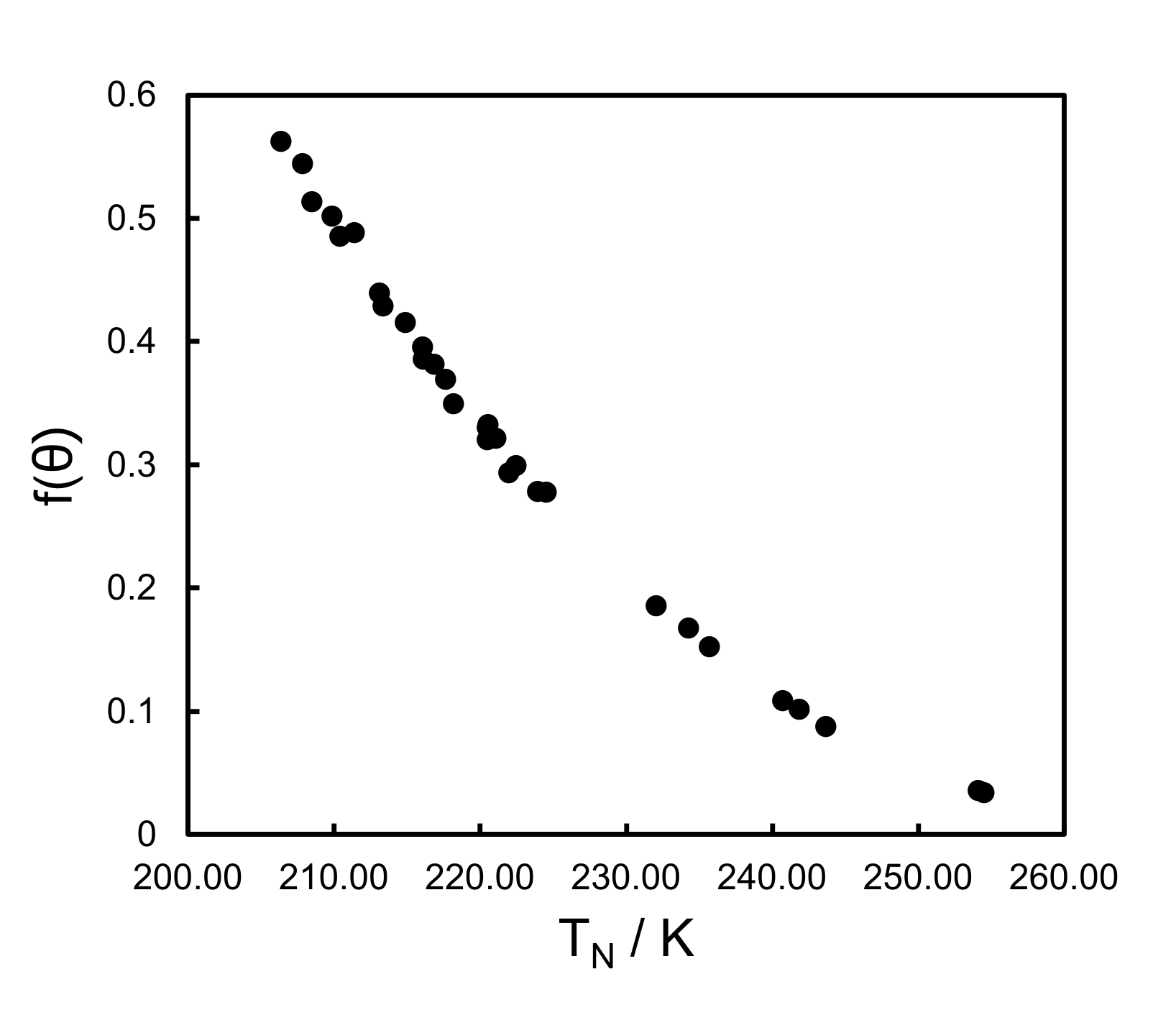} b)
    \includegraphics[clip,scale=0.19,angle=0.0]{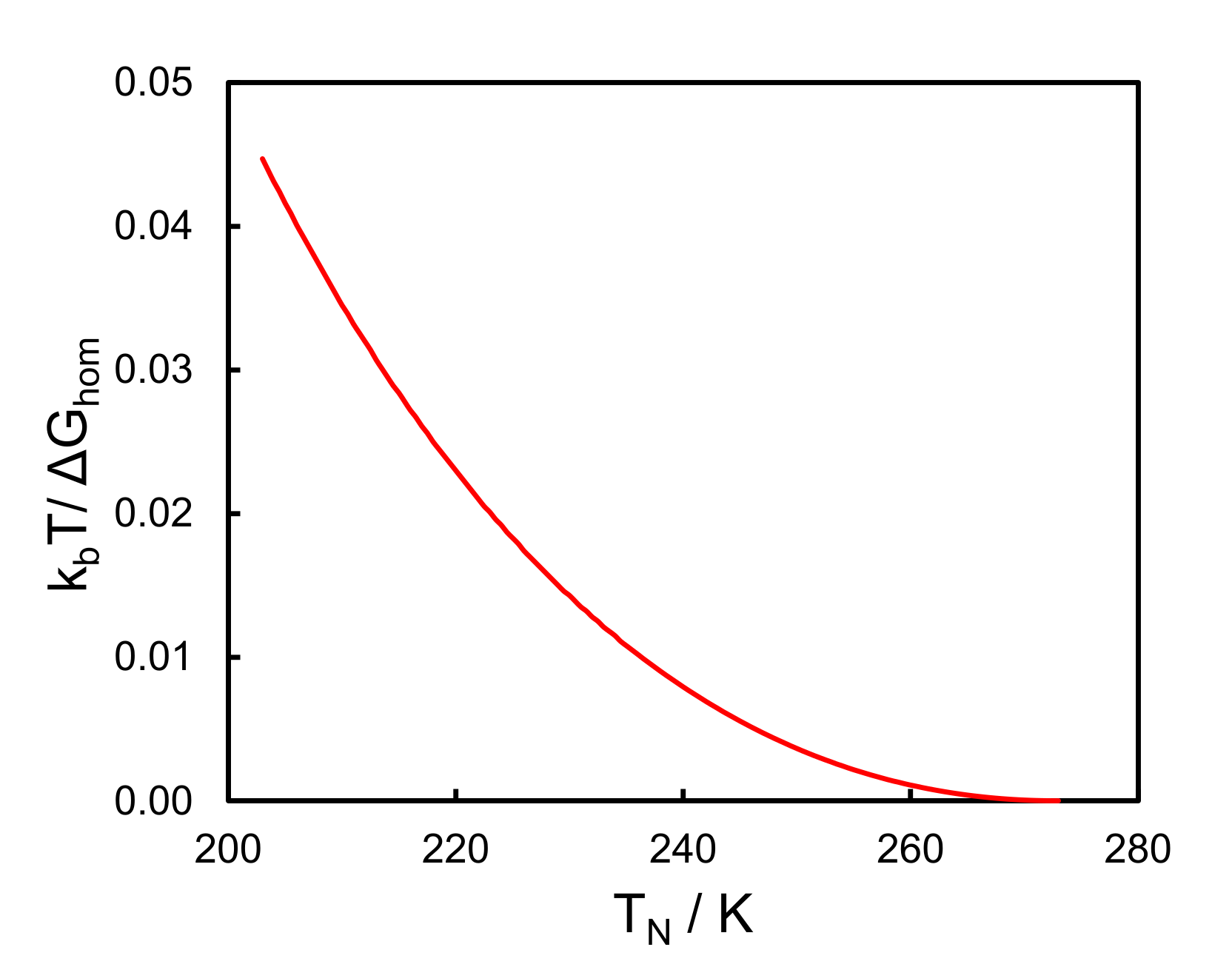} c)
    \caption{(a) Contact angle versus the nucleation temperature for different substrates (see lengend). 
    \textcolor{black}{For the ice-like substrates, re-analysed here from our previous work \cite{camarillo2024effect}, the legend indicates the exposed
    plane (secondary prismatic in all cases) and the percentage by with the unit cell was either expanded (exp) or compressed (comp).}
    (b) $f(\theta)$ versus
    the nucleation temperature for all substrates.  (c) Inverse homogeneous nucleation barrier (in $k_BT$ units) versus temperature.}
    \label{fig:thetavsTn}
\end{figure}

Once we compute the nucleation rate, we can also estimate 
nucleation barriers and contact angles as explained in Sec. \ref{sec:contact}. 
This task requires a comprehensive control of the parameters that govern homogeneous
nucleation, which we have from our previous studies \cite{espinosa2016interfacial,espinosa2016time}.
The first step toward the calculation of $\Delta G_{het}^c$
is to obtain the kinetic pre-factor
from the CNT rate expression given in Eq. \ref{eq:jhetcnt}. 
The kinetic pre-factor, $A$, contains three contributions, the number density 
of nucleating sites, $\rho_s$, the attachment rate, $f^+_{het}$, and
the Zeldovich factor, $Z_{het}$. 

The first factor, $\rho_s$, is simply given by the number density 
of molecules in the substrate plane exposed to the liquid. We report
$\rho_s$ in Table \ref{tab:rate-results}. 
$\rho_s$ varies from one substrate to another, but it is in all 
cases of the order of 10 nm$^{-2}$. 

$f^+_{het}$ and $Z_{het}$ are obtained from our 
previous work on homogeneous nucleation
\cite{espinosa2016interfacial,espinosa2016time} using Eqs. \ref{eq:f+} and \ref{eq:Z} respectively.
According to these equations, in order to obtain 
$f^+_{het}$ and $Z_{het}$, one first needs $\theta$, which, in turn, 
depends on $f^+_{het}$ and $Z_{het}$ (see Section \ref{sec:contact}). 
The solution is to iteratively solve the equations leading to $\theta$, $f^+_{het}$ and $Z_{het}$. 
Initially, $\Delta G^c_{het}$, is obtained from 
Eq. \ref{eq:jhetcnt} assuming an initial value for $f(\theta)$ between 0 and 1
to obtain $Z_{het}$ and $f^+_{het}$ via Eqs. \ref{eq:f+} and \ref{eq:Z}
respectively. $f(\theta)$ is then obtained via Eq. \ref{eq:deltaGhomhet} and
$Z_{het}$ and $f^+_{het}$ are accordingly updated with the new 
$f(\theta)$ value. With these new values of the attachment
rate and the Zeldovich factor $\Delta G^c_{het}$ is computed again via Eq. \ref{eq:jhetcnt}, 
giving rise to a new value of $f(\theta)$. This process is repeated until convergence, 
which is quickly reached as shown for one of the 
systems under study in Figure. \ref{fig:iterations}, where we represent $f(\theta)$ versus the
iteration number for two different initial guess values for $f(\theta)$.

The attachment rate and the Zeldovich factor for the substrates under study
are reported in Table \ref{tab:rate-results}.
The heterogeneous attachment rate 
is of the same order as $f^+_{hom}$ 
($\sim$ 10$^{13}$s$^{-1}$)\cite{espinosa2016interfacial,espinosa2016time}, 
although somewhat lower because
$h(\theta)$ in Eq. \ref{eq:f+} is a fraction smaller than 1
(recall that $h(\theta)$ is the ratio between the contact areas with liquid water of the 
homogeneous and the heterogeneous nuclei).
Also, $Z_{het}$ is of the same order as $Z_{hom}$ ($\sim$ 10$^{-2}$) \cite{espinosa2016interfacial,espinosa2016time}, although
in this case $Z_{het}$ is larger than $Z_{hom}$, which can be easily understood
by taking into account that $f(\theta)$ is a fraction smaller than 1 in Eq. \ref{eq:Z}.
At the end of the day, the product of $f^+_{het}$ and $Z_{het}$ is 
very close to that corresponding to homogeneous nucleation given that 
the attachment rate is larger in homogeneous than in heterogeneous nucleation, but 
the opposite is true for the Zeldovich factor.

It is also worth realizing that $A$ is rather insensitive to temperature changes because
$\rho_s$ is constant, $f^+$ increases as temperature increases due to the faster diffussivity
of the liquid, but $Z$ decreases more or less to the same extent due to the fact that the free energy
barrier increases as temperature goes up and decreases its curvature (recall that $Z$ 
is proportional to the square root of the curvature of the free energy barrier at the maximum). 

The nucleation free energy barrier heights for heterogeneous nucleation, $\Delta G^c_{het}$, 
obtained from the calculations of $J_{het}$ as previously described (in an iterative 
process), are reported
in Table \ref{tab:rate-results}. All barriers are about 15 $k_BT$ high. This is not a coincidence,
but rather a direct consequence of the fact that we are comparing different substrates 
for a particular value of the rate ($J_{het}=10^{24}$ m$^{-3}$s$^{-1}$). 
Since the kinetic pre-factor in Eq.\ref{eq:jhetcnt} does not substantially vary with 
either temperature or substrate type, it follows that the nucleation barrier does not change much 
from one substrate to another either (for a fixed rate). 
In other words, the rates accessible to spontaneous nucleation
entail free energy barriers of about 15 $k_BT$, which is a reasonable value for a 
spontaneous stochastic nucleation process \cite{filion2010crystal}. 

What does change from one substrate to another is the temperature at which 
spontaneous nucleation is achieved (discussed in Section \ref{sec:rate_Tn}) 
and the contact angle, $\theta$, which we discuss now. 
The contact angle, defined in Fig. \ref{fig:contact-angle}, is a useful parameter to quantify the ability 
of a substrate to induce heterogeneous nucleation. 
It varies from 180º, when nucleation is not induced by the substrate (homogeneous nucleation limit), 
to 0º, when the substrate is fully wetted by the nucleating phase, that grows without surmounting any free
energy barrier. 

We report the contact angles 
in Table \ref{tab:rate-results}. 
Obviously, the lower is the contact angle, the better ice wets the substrate and the higher is the
nucleation temperature. 
Contact angles for the most efficient nucleants under consideration, the stretched/compressed ice substrates,  are in the range of 40-55º 
whereas those of the least efficient one, the bcc-100 substrate, lie in the range of 85-95º. 
We plot in Fig. \ref{fig:thetavsTn}(a) the contact angles of all substrates investigated. 
There seems to be a rather linear dependence of the contact angle with the nucleation temperature. 
We include in green in the figure the point $\theta=0$ for the melting temperature of the model, 273 K, indicating
that a substrate for which the contact angle would be 0 (full wetting) would cause crystallization at the 
melting temperature. The trend of the data obtained at lower temperatures from simulations seems to be consistent with this idealised point. 

The seemingly linear dependence of $\theta$ with temperature can be understood as follows. 
First of all, $f(\theta)$ is a sigmoid-like function with the inflection point at $\pi/2$. 
Therefore, for $\theta$ values around $\pi/2$, $f(\theta)$ looks linear. Being $\theta$
and $f(\theta)$ linearly related to each other in a certain range of $\theta$'s, a linear
$f(\theta)$ versus $T_n$ dependence is expected too. This is in fact what is seen in Fig. 
\ref{fig:thetavsTn}(b) (understandably, for high temperatures, that correspond to contact angles away from $\pi/2$, the plot deviates from linearity).
We now focus on $f(\theta)$ to understand why it looks linear. $f(\theta)$ is the ratio 
between the heterogeneous and the homogeneous nucleation barriers. The former is 
about 15$k_BT$ for all substrates for the reasons previously discussed. Therefore, 
the $f(\theta)$ dependence with temperature is dictated by that of the inverse homogeneous 
nucleation barrier, which we represent in \ref{fig:thetavsTn}(c) and, indeed, has the same 
shape as $f(\theta)$ and shows a nearly linear dependency for low temperatures ($\theta$'s close to $\pi/2$).

\begin{figure}[H]
    \centering
    \includegraphics[clip,scale=0.20,angle=0.0]{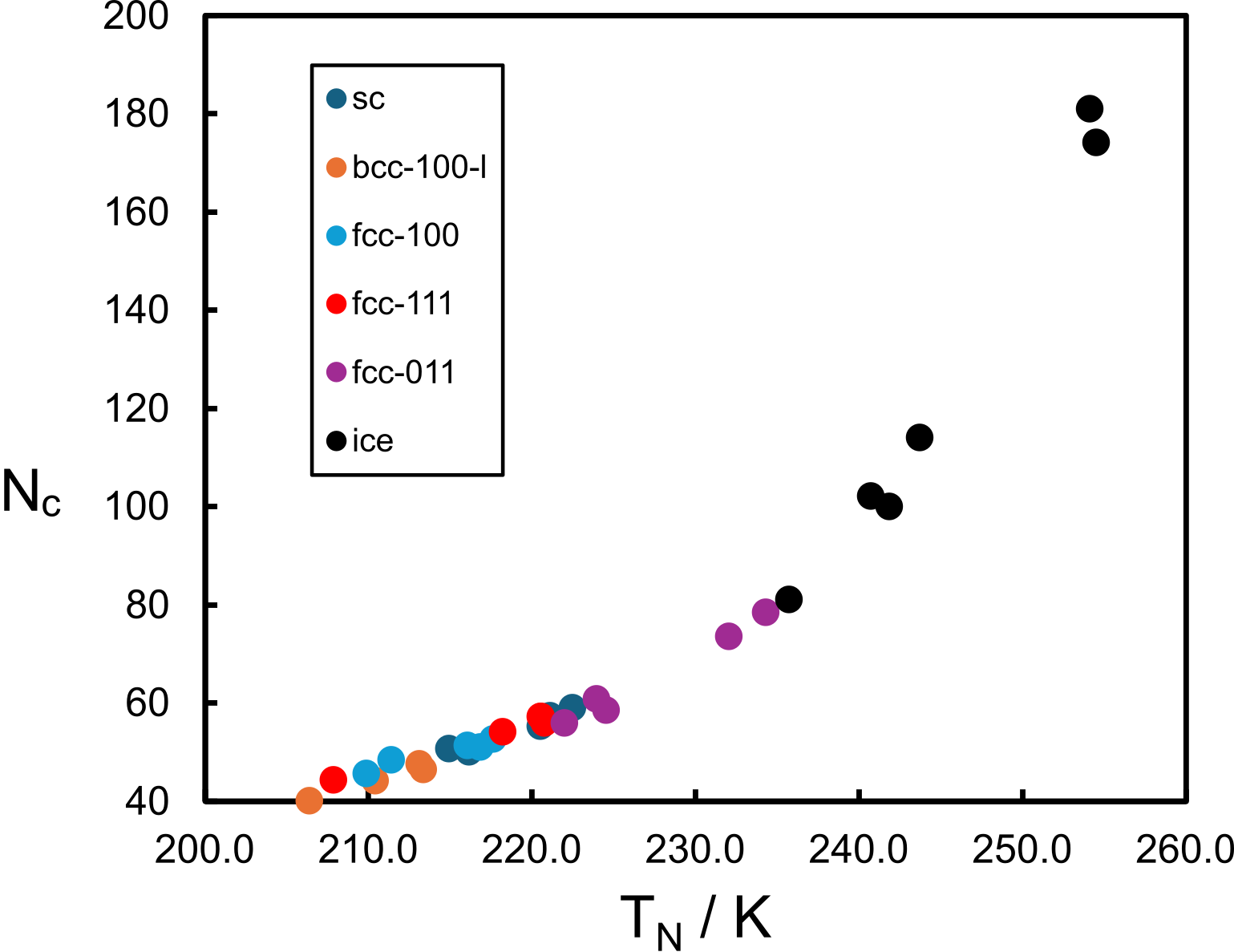}
        \caption{Number of particles in the critical nucleus 
        for different types of substrate (see legend) versus the
        nucleation temperature.}
    \label{fig:ncvstn}
\end{figure}

Having $f(\theta)$ we can estimate also the number of particles in the critical cluster in heterogeneous nucleation using $N_{c,het}=N_{c,hom}f(\theta)$.  We take $N_{c,hom}$ from our previous work on homogeneous nucleation \cite{espinosa2014homogeneous,espinosa2016time,espinosa2016interfacial}. We report in Table \ref{tab:rate-results} $N_{c,het}$ for all systems under study and in Fig. \ref{fig:ncvstn}  
we plot $N_{c,het}$ versus the nucleation temperature for all substrates investigated in this work, alongside the
ice-like substrates from our previous paper on heterogeneous ice nucleation \cite{camarillo2024effect}.
Critical cluster sizes in our spontaneous nucleation trajectories vary in the range of 50 to nearly 200 particles. 
The better ice nucleant is a substrate, the higher is $T_n$ and the larger is $N_{c,het}$. This can be understood using
CNT, that predicts the size of the critical cluster to be equal to $2 \Delta G^c_{het}/|\Delta \mu|$. Since, as previously discussed, 
$\Delta G^c_{het}$ is constant (about 15 $k_BT$) the size goes down as the nucleation temperature goes down too because
the chemical potential difference between ice and water increases upon cooling.  In other words, good ice nucleants are capable of nucleating larger critical clusters at higher temperatures. 
Interestingly, in Section \ref{sec:critnucsize} we show that the critical cluster sizes predicted with CNT are consistent with a microscopic analysis of the simulation nucleation trajectories. 

Our approach enables determining the contact angle without 
 a microscopic analysis of 
simulation configurations, which is not a trivial task. In fact, several algorithms have been 
developed for such purpose and there is not a standard procedure yet \cite{vuckovac2019uncertainties,jiang2019recent,santiso2013calculation,wang2022contactanglecalculator}.
In solid-liquid nucleation the situation is particularly delicate because the interface
is not as clearly defined as in vapor-liquid nucleation due to the similarity in density
between both phases. 
As an example, an attempt to obtain the solid-liquid interfacial free energy for NaCl
using Young's equation and a contact angle estimate from a contour of the nucleus
yields $\sim$ 40 mJ/m$^2$, \citep{zykova2005alkali,zykova2005physics}, which is in strong disagreement
with the $\sim$ 100  mJ/m$^2$ obtained with different techniques that do not rely on contact angle estimates \cite{benet2015interfacial,bahadur2007surface,espinosa2015crystal}.

\subsection{Structural and microscopic analysis}

\subsubsection{Induced structure of the liquid}
\begin{figure}[H]
    \centering
    \includegraphics[clip,scale=0.20,angle=0.0]{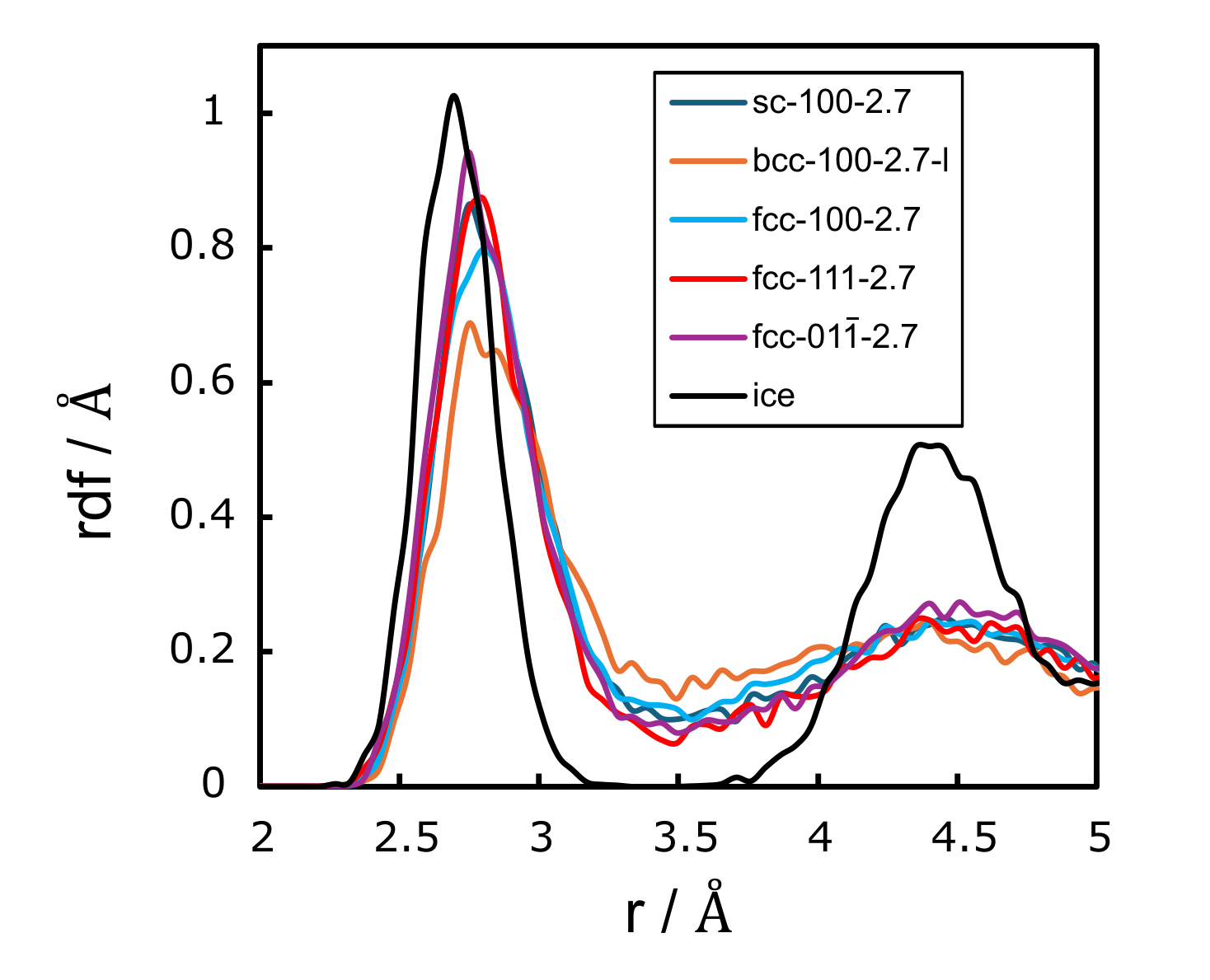}(a)
    \includegraphics[clip,scale=0.20,angle=0.0]{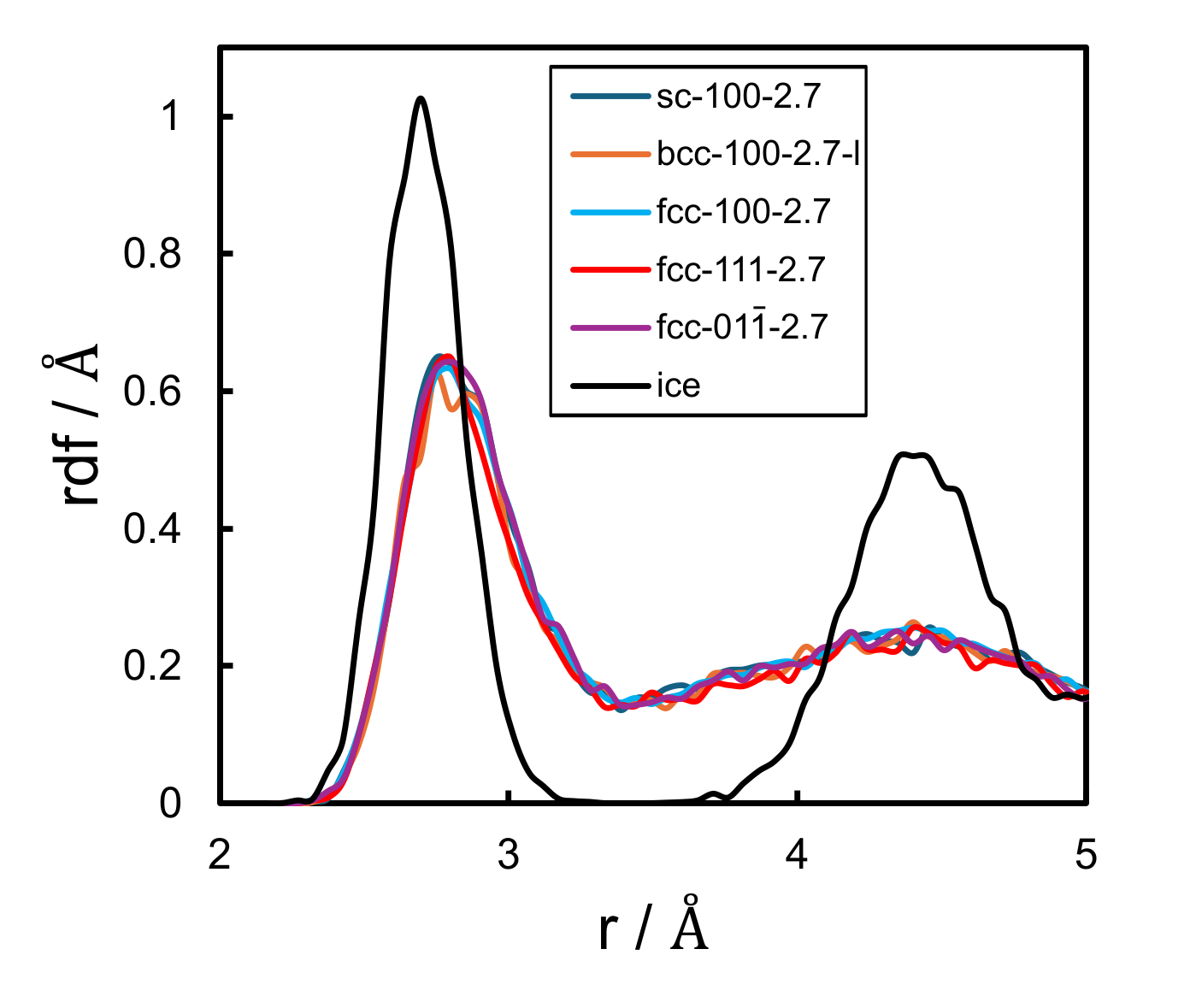}(b)
    \caption{(a) Radial distribution function of a 4\AA~ thick liquid slab adjacent to the substrate at 235 K. Different colours correspond to different substrates (see legend). The black curve corresponds to a slab of the same width of bulk ice. (b): Same as (a) but considering a liquid slab far away from the substrate.}
    \label{fig:rdf}
\end{figure}

To understand at a more microscopic level the ice nucleating abilities of the different substrates 
we evaluated the rdf of 4 \AA~ thick slabs of the liquid immediately adjacent to the substrate at a given temperature (235 K).
The comparison is established for a given nnd (2.7 \AA). 
No nucleation occurred during the time interval in which the rdf's were obtained. We focus on the structure of the liquid layer induced by the 
substrate. 
The result is shown in Fig. \ref{fig:rdf}(a) for all substrates under study (see legend for the color code). 
We include, for comparison, the rdf of a bulk ice slab of the same width (in black).
The rdfs are normalised so that they integrate to the same area at the end of the range in which they were calculated (they do not tend
to 1 as in a bulk liquid because we are only focusing on the particles contained in a certain slab). Therefore, the absolute height of
the peaks does not have any physical meaning and we just focus on the relative differences. We observe that there is a correlation 
between ice nucleating ability and height of the rdf first peak. 
The best ice nucleant (among non ice-like substrates), the fcc-01$\bar{1}$ substrate, has the highest first rdf peak (in purple) whereas the worst one, bcc-100-l,
induces the lowest first rdf peak (in orange).
To test that the differences between rdf curves are not due to statistical noise, we compute the rdf in a liquid slab of the same width 
but far away from the substrate. This is shown in Fig. \ref{fig:rdf}(b). As expected, the rdf functions for all substrates are the same within noise. 
If one compares the rdf first peak height from the bulk liquid slab with that of the slab adjacent to the substrate, it is evident 
that the latter is significantly higher (as much as about 50\% higher for the case of fcc-01$\bar{1}$, the most efficient substrate). 
It is therefore clear that the substrates enhance the structure of the adjacent fluid and that the more structured is the fluid the higher
is the temperature at which ice nucleation is observed. \textcolor{black}{That local water structure correlates with ice nucleation ability has been noted in several earlier papers (see e. g. Ref. \cite{soniJCP2024machinelearning} and references therein)}.

\textcolor{black}{The rdf results in Fig. \ref{fig:rdf}(a) confirm that, contrary to intuition, the
fcc-01$\bar{1}$ orientation induces a more structured liquid than the fcc-111, despite lacking hexagonal symmetry. To confirm this point,
we compare the average $Q_6$ order parameter \cite{steinhardt1983bond} for the liquid molecules contained in a 4 \AA~ thick slab
adjacent to the substrate at 235 K. A 4 \AA~ cutoff was used for the calculation of $Q_6$. We obtain
average $Q_6$ values of 0.376(1) and 0.381(1) for the fcc-111 and fcc-01$\bar{1}$ lattices, respectively.
As a reference, a 4 \AA~ slab in the liquid far from the substrate has an average $Q_6$ of 0.371(1).
This result indicates a higher degree of order in the liquid molecules adjacent to the fcc-01$\bar{1}$ substrate, which is
consistent with the higher rdf and the higher nucleation temperature of this substrate compared
to the fcc-111 orientation.}

\subsubsection{Critical nucleus}
\label{sec:critnucsize}
\begin{figure}[H]
    \centering
    \includegraphics[clip,scale=1.2,angle=0.0]{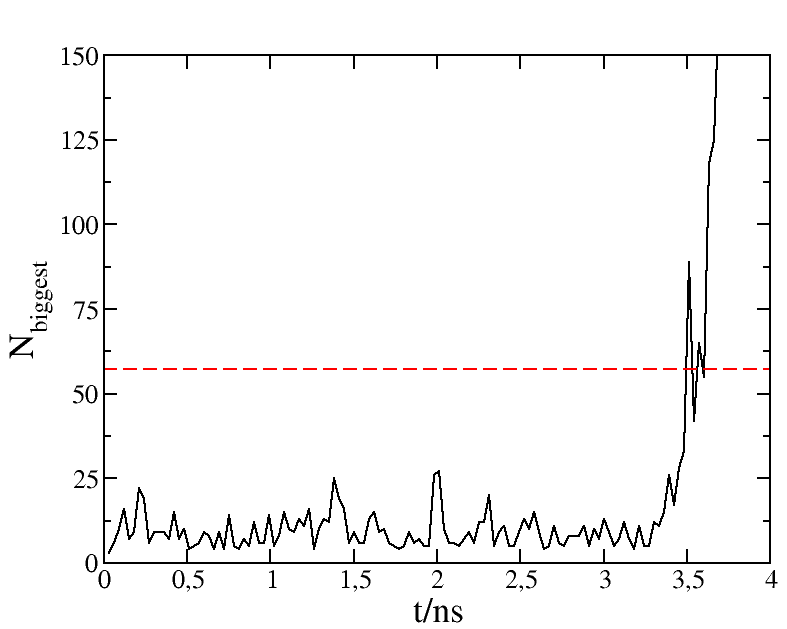}(a)
    \includegraphics[clip,scale=0.40,angle=0.0]{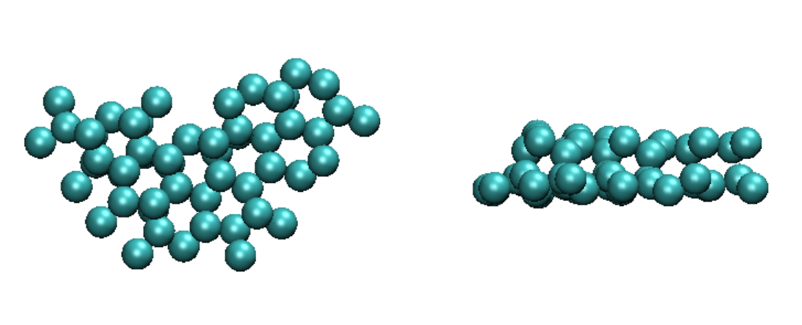}(b)
    \caption{(a) Number of particles in the biggest ice cluster versus time in a nucleation trajectory on the sc-100-2.7 substrate at 221 K. 
A particle is labelled as ice-like if its $\bar{q}_6$ is larger than 0.39. 
Neighbors within 3.51 \AA~ are considered to compute $\bar{q}_6$ for each molecule.  The same distance is employed to identify molecules belonging to the same cluster.  (b) Snapshot
    of a cluster identified as critical. Left, top view (basal plane), right, side view (secondary prismatic plane).}
    \label{fig:nbiggest}
\end{figure}

In Table \ref{tab:rate-results}
and in Fig. \ref{fig:ncvstn} we provide the critical cluster sizes 
obtained by analysing our nucleation rate data with CNT. 
Here, we test the consistency of the results provided by this thermodynamic theory with a microscopic
analysis of the spontaneous nucleation trajectories. In particular, we analyse the number of particles in 
the biggest ice cluster along time and try to identify the critical cluster from such an analysis. To calculate 
the number of particles in the 
biggest ice cluster we use the $\bar{q}_6$ order parameter proposed in Ref. \cite{lechner2008accurate} tuned for the identification
of ice-like molecules and clusters as described in Ref. \cite{espinosa2016seeding}. 
In Fig. \ref{fig:nbiggest}(a)
we plot the number of particles in the biggest ice cluster, $N_{biggest}$, versus time for a nucleation trajectory on an 
sc-100-2.7 substrate at 221 K. 
Some small fluctuations in $N_{biggest}$ are observed until at 3.5 ns there is a large one that leads to the 
formation of a big cluster. Its size fluctuates for about 0.2 ns and it then irreversibly grows. We identify such fluctuations previous to irreversible growth with the critical cluster wandering around the free energy barrier top. 
According to our CNT analysis, the critical cluster contains 57 molecules (see Table \ref{tab:rate-results}), which is 
indicated with a horizontal red dashed line in the figure.  Such size is perfectly consistent with the 
critical cluster size fluctuations identified with the microscopic analysis of the simulation trajectory.

It is perhaps surprising that a thermodynamic theory such as CNT provides microscopically consistent 
results down to such small clusters. From our experience in ice \cite{espinosa2018homogeneous}, hydrate \cite{zeron2025homogeneous} and NaCl \cite{lamas2021homogeneous} crystallization and 
on water condensation and cavitation \cite{camarillo2025condensation} it is not the fist time that CNT performs well in the limit 
of very small nuclei. 
Given the agreement between CNT and our microscopic determinations of the critical cluster size, it is worth noting that the line‐tension term in the heterogeneous nucleation CNT framework —omitted here for simplicity in order to extract nucleation parameters by comparing homogeneous and heterogeneous nucleation rates—appears to be negligible.

\textcolor{black}{It is worth realising that in our previous Seeding estimates of the homogeneous 
nucleation rate \cite{espinosa2016time} we did not use the capillarity approximation, which
considers that $\gamma_{il}$ is fixed at the ice-liquid water coexistence value. Instead, we allowed $\gamma_{il}$ 
to change with temperature, while keeping the mathematical form of the CNT expressions. With this approach, we found that 
$\gamma_{il}$ decreases as water is supercooled. Had we used $\gamma_{il}$ at coexistence we would 
have obtained larger nucleation barriers ($\Delta G_c \propto \gamma_{il}^3$) and, consequently, larger critical nucleus sizes
($N_c \propto \Delta G_c$). In the particular case
of the analysis of nucleation on an sc-100-2.7 substrate at 221 K, 
had we used $\gamma_{il}$ at coexistence (35 mN/m) instead
of $\gamma_{il}$ at 221 K (28 mN/m) \cite{espinosa2016time}, we would have predicted with CNT a critical cluster
containing almost twice as many particles (112 versus 57). Therefore, the agreement between the microscopic analysis
of the Molecular Dynamics trajectory 
and CNT shown in Fig. \ref{fig:nbiggest} would have been deteriorated had the capillarity
approximation been considered. Such an agreement could have also been worsened if a different order parameter
to identify the number of particles in the critical cluster had been employed (different order parameters give different number of molecules in the cluster for the same configuration \cite{filion2010crystal}). In this respect, it is
important to stress that the order parameter
we used to produce Fig. \ref{fig:nbiggest} was not tuned a posteriori to give a good agreement with CNT, but a priori using the mislabelling 
criterion proposed in Ref. \cite{espinosa2016seeding}, which seeks balancing the number of wrongly labelled particles in 
both bulk phases.}

\begin{figure}[H]
    \centering
    \includegraphics[clip,scale=0.40,angle=0.0]{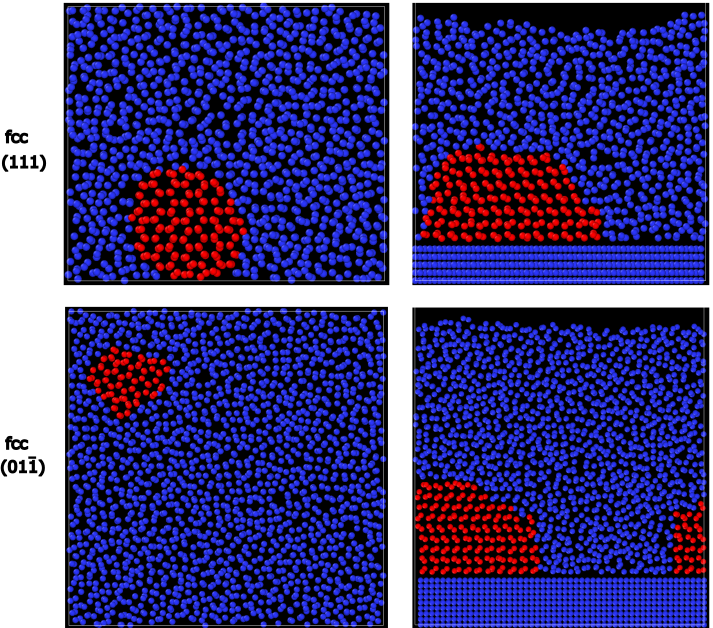}
    
    \caption{\textcolor{black}{
    Left panels show a 6 \AA-thick slab parallel and adjacent to the substrate containing a freshly nucleated ice cluster (in red), where hexagons typical of the basal plane are clearly visible. Right panels show a 6 \AA-thick slab perpendicular to the substrate, passing through the cluster’s center of mass and exposing the secondary prismatic plane. Stacking faults are evident from the absence of a long-range repetitive pattern along the direction normal to the substrate. The top and bottom rows correspond to fcc-111 and fcc-100 substrates, respectively.}}
    \label{fig:redcluster}
\end{figure}

In Fig. \ref{fig:nbiggest}(b) we show a top and a side view of one configuration of the critical cluster.
Of course, it is difficult to obtain an average size and shape from individual configurations, particularly
for such small sizes. However, one can already identify the characteristic hexagons of the basal plane in 
the top view and the secondary prismatic pattern in the side view. It seems, then, that the basal plane preferentially nucleates on top of an sc-100 substrate. 
\textcolor{black}{In the left panels of Fig. \ref{fig:redcluster} we show recently nucleated clusters on an fcc-111 (top) and on an
fcc-100 (bottom) substrate. Again, typical hexagons of the basal plane are clearly visible.}
We have checked that the same is true for
all other substrates studied in this work (bcc-100 and fcc-01$\bar{1}$). 
Moreover, we have observed that when the cluster grows 
there are staking faults and the growth phase is neither ice Ih nor ice Ic, but a stacking hybrid between both \cite{malkin2015stacking,kuhs2012extent,lupi2017role,quigley2014communication}.
\textcolor{black}{Snapshots of growing clusters with stacking faults are shown in the right panels of Fig. \ref{fig:redcluster}}.

\section{Conclusions}
We use the mW water model to investigate the 
ability of different model lattices to nucleate ice from 
supercooled water. 
In this work we investigate simple cubic, body centred cubic
and face centred cubic lattices exposing the 100 orientation 
to the liquid. In the case of the face centred cubic lattice
we also expose the 111 and the 01$\bar{1}$ orientations.
We compare the results obtained in this work for these substrates
with those from Ref. \cite{camarillo2024effect}, where we studied
substrates with a stretched/compressed ice structure. 
To focus exclusively on the role of the 3D lattice arrangement and
orientation, the substrates are composed of water molecules which are fixed at their
lattice positions. 
We run simulations at low temperatures
and 0 bar pressure and wait for heterogeneous nucleation 
to spontaneously occur. We obtain the heterogeneous nucleation 
rate from such simulations, which enables us to rank 
the ice nucleation ability of the substrates investigated. 
By comparing heterogeneous and homogeneous nucleation rates under
the framework of Classical Nucleation Theory, we obtain 
many relevant parameters describing the thermodynamics and the kinetics
of the nucleation process. 
We also perform a microscopic analysis of the substrate-liquid interface
and of the nucleation trajectories to gain a better understanding of
heterogeneous nucleation. 
These are the main conclusions we draw from our investigation:

\begin{itemize}

\item fcc substrates exposing the 01$\bar{1}$ orientation are the most effective ice nucleants
among all model lattices investigated in this work, causing nucleation up to 25 K above the poorest
nucleant, the bcc lattice. 

    \item 
The effect of the lattice parameters on the ice nucleating ability is strong 
for ice-like lattices, 
moderate for the fcc-01$\bar{1}$ substrate and quite mild for the rest of the studied substrates. Consequently, the better is an ice nucleant the more sensitive is the 
nucleation efficiency to its lattice parameters. 

\item
Even for mW, that is a short-range interaction water model, there is a noticeable influence
on the nucleation ability of crystal planes underneath the one which is in direct contact with the liquid. 
Thus, fcc-100, sc-100 and bcc-100 substrates, all of them exposing a quadrangular 2D lattice to the 
liquid, present clear differences in the temperatures at which they induce ice nucleation (up to $\sim$
15 K difference in nucleation temperature between the best and the worst nucleants). 

\item
There is a strong impact of the lattice orientation on the ice nucleation temperature. 
For instance, there is a $\sim$ 25 K difference in nucleation temperature between the most (01$\bar{1}$) and the least (100) efficient studied orientations of
the fcc lattice.

\item
There is a non-trivial correlation between the symmetry of the exposed lattice plane and 
the nucleation temperature. In principle, due to the hexagonal symmetry of the ice lattice, 
one would expect that hexagonal planes are more efficient ice nucleants. 
However, the 111 orientation of the fcc lattice, which is a closed-packed hexagonal
2D lattice, nucleates ice less efficiently
(at about 15 K lower temperature) than the 01$\bar{1}$ plane, whose symmetry is not hexagonal. 

\item
An analysis of the radial distribution of the liquid layer adjacent to the substrate confirms that 
better ice nucleants enhance more strongly the liquid structure.  

\item
By comparing heterogeneous with homogeneous nucleation rates, and using the Classical Nucleation formalism,
we put forward a methodology to obtain contact angles, nucleation barriers, kinetic pre-factors, and critical cluster sizes 
in heterogeneous nucleation. These parameters, which are very useful to characterise the nucleation 
process, are obtained through an iterative algorithm that converges quite quickly. Contact angles of about 90$^o$ are obtained
for the poorest nucleant (bcc) and of $\sim$ 45$^o$ for the ice-like substrates, which are the best nucleants. 

\item
We find consistency between the critical cluster size obtained from our CNT-based analysis of the 
computed nucleation rates and that inferred from a microscopic analysis of the nucleation trajectories. 
Such consistency supports the validity of CNT to study heterogeneous nucleation at deep supercooling, where
small clusters nucleate, 
and suggests that line tension effects (which we ignore in our analysis) are not important.

\item 
We find that the basal plane of ice is the one that nucleates on all substrates, highlighting the
facility with which this plane assembles as compared to other ice planes. Moreover, 
we find that ice nuclei grow with stacking faults.

\end{itemize}

\textcolor{black}{Given that the studied substrates are generic lattices, the main aim of our work is to gain qualitative insight into the effect of lattice structure on ice-nucleating ability. Nevertheless, our approach can be gradually made more complex to provide a closer representation of real substrates. For instance, interactions can be modified at will for a specific lattice to isolate the effect of interactions alone and to gradually make substrate-liquid interactions more
realistic. Additionally, one can manipulate the smoothness or the order of the interface, the chemical heterogeneity of the surface, or the crystallinity of the substrate to gain qualitative understanding on how these factors,
that play a role in real substrates, influence ice nucleation.}

\section{Data Availability Statement}
The data that supports the findings of this study are available within the article.

\begin{acknowledgments}
This work was funded by Grants No. 
PID2022-136919NB-C31 and PID2022-136919NB-C32 of the MICINN. The authors gratefully acknowledge the Universidad Politecnica de Madrid (www.upm.es) for providing computing resources on Magerit Supercomputer.
E. S. thanks
Carlos Vega for his guidance in pilgrimages to remote hermitages on sunny Fridays. 
M.M.C. wishes to express her deepest gratitude to Carlos Vega, her Ph.D. supervisor. He awakened her scientific curiosity through his captivating stories about water. His enthusiasm and eloquence opened the door to the world of science for her, and his words —always full of wisdom— have stayed with her ever since. As he always reminds her: ``we’ve fought in worse arenas''.

\end{acknowledgments}

\clearpage
\section{REFERENCES}

\clearpage

\end{document}